\documentclass[table]{article}
\usepackage[utf8]{inputenc}
\usepackage{amsmath}
\usepackage{amsfonts}
\usepackage{physics}
\usepackage{amsmath}
\usepackage{tikz}
\usepackage{mathdots}
\usepackage{yhmath}
\usepackage{amsmath}
\usepackage{cancel}
\usepackage{hyperref}
\usepackage{siunitx}
\usepackage{array}
\usepackage{subfig}
\usepackage{multirow}
\usepackage{svg}
\usepackage{enumitem}
\usepackage{booktabs}
\usepackage{amssymb}
\usepackage{xfrac}
\usepackage{nicefrac}
\usepackage{svg}
\usepackage{gensymb}
\usepackage{physics}
\usepackage{amsmath}
\usepackage{tikz}
\usepackage{mathdots}
\usepackage{yhmath}
\usepackage{cancel}
\usepackage{siunitx}
\usepackage{array}
\usepackage{multirow}
\usepackage{amssymb}
\usepackage{gensymb}
\usepackage{tabularx}
\usepackage{extarrows}
\usepackage{booktabs}
\usetikzlibrary{fadings}
\usetikzlibrary{patterns}
\usetikzlibrary{shadows.blur}
\usetikzlibrary{shapes}
\usepackage{mathdots}
\usepackage{yhmath}
\usepackage{cancel}
\usepackage{siunitx}
\usepackage{array}
\usepackage{multirow}
\usepackage{amssymb}
\usepackage{gensymb}
\usepackage{tabularx}
\usepackage{extarrows}
\usepackage{booktabs}
\usetikzlibrary{fadings}
\usetikzlibrary{patterns}
\usetikzlibrary{shadows.blur}
\usetikzlibrary{shapes}
\usepackage{physics}
\usepackage{amsmath}
\usepackage{tikz}
\usepackage{mathdots}
\usepackage{yhmath}
\usepackage{cancel}
\usepackage{siunitx}
\usepackage{array}
\usepackage{multirow}
\usepackage{amssymb}
\usepackage{gensymb}
\usepackage{tabularx}
\usepackage{extarrows}
\usepackage{csquotes}
\usepackage{booktabs}
\usetikzlibrary{fadings}
\usetikzlibrary{patterns}
\usetikzlibrary{shadows.blur}
\usetikzlibrary{shapes}
\usepackage{caption}
\usepackage{geometry}
\usepackage{booktabs}
\usetikzlibrary{fadings}
\usetikzlibrary{patterns}
\usetikzlibrary{shadows.blur}
\usetikzlibrary{shapes}
\usepackage{tabulary}
\usepackage{booktabs}
\usepackage{sectsty}
\definecolor{ForestGreen}{RGB}{34,139,34}
\definecolor{MidnightBlue}{RGB}{25, 25, 112}
\usepackage{dcolumn}
\newcolumntype{d}[1]{D..{#1}}

\hypersetup{
    colorlinks=true,
    urlcolor=MidnightBlue,
    linkcolor=ForestGreen,
    citecolor=ForestGreen
}

\usepackage[nottoc]{tocbibind}

\usepackage{dcolumn,booktabs}
\newcolumntype{d}[1]{D{.}{.}{#1}}

\newcommand\Underset[2]{\underset{\textstyle #1}{#2}}
\newcommand{\signnn}{\hspace{-0.125cm}$^{***}$} 
\newcommand{\signn}{\hspace{-0.125cm}$^{**\ }$} 
\newcommand{\sign}{\hspace{-0.125cm}$^{*\ \ }$} 
\newcommand{\nosign}{\hspace{-0.125cm}$^{\ \ \ }$}

\sectionfont{\fontsize{10}{15}\selectfont}

\usepackage{amsthm,amssymb}
\renewcommand\qedsymbol{$\blacksquare$}

\sectionfont{\large}
\subsectionfont{\normalsize}
\subsubsectionfont{\normalsize}

\newcommand\blfootnote[1]{%
  \begingroup
  \renewcommand\thefootnote{}\footnote{#1}%
  \addtocounter{footnote}{-1}%
  \endgroup
}

\usepackage[font={small}]{caption}

\usepackage[
backend=biber,
style= authoryear,
maxcitenames = 2,
uniquename=false,
uniquelist=false,
maxbibnames=9
]{biblatex}
\bibliography{bib.bib}

\author{
  Tamay Besiroglu\footnotemark[1]\\
       \small{MIT FutureTech}
  \and
  Nicholas Emery-Xu\footnotemark[1]\\
    \small{UCLA Dept. of Economics, MIT FutureTech}
   \and
   Neil Thompson\footnotemark[2]\\
       \small{MIT FutureTech}
}
\date{}

\title{Economic impacts of AI-augmented R\&D}

\begin{document}
\maketitle

\begin{abstract}
\noindent Since its emergence around 2010, deep learning has rapidly become the most important technique in Artificial Intelligence (AI), producing an array of scientific firsts in areas as diverse as protein folding, drug discovery, integrated chip design, and weather prediction. As more scientists and engineers adopt deep learning, it is important to consider what effect widespread deployment would have on scientific progress and, ultimately, economic growth. We assess this impact by estimating the idea production function for AI in two computer vision tasks that are considered key test-beds for deep learning and show that AI idea production is notably more capital-intensive than traditional R\&D. Because increasing the capital-intensity of R\&D accelerates the investments that make scientists and engineers more productive, our work suggests that AI-augmented R\&D has the potential to speed up technological change and economic growth.
\end{abstract}

\blfootnote{* Joint first authors}
\blfootnote{$\dagger$ Corresponding author \href{mailto:neil_t@mit.edu}{neil\_t@mit.edu}}
\blfootnote{For helpful comments we are grateful to Philip Trammell, Basil Halperin, Matt Clancy, Charlotte Siegmann, Ege Erdil, Gabriel Filipe, Wensu Li, Tom Davidson, Jaime Sevilla, and Nur Ahmed.}
\blfootnote{The authors are grateful  to Open Philanthropy for financial support. Nicholas Emery-Xu is also grateful to the UCLA Graduate Division and the Future of Humanity Institute for financial support.}

\section{Introduction}
\label{sec:Introduction}

In this paper, we consider what effect the adoption of Artificial Intelligence (AI) within science and engineering will have on idea production and, subsequently, on productivity and economic growth. Unlike previous work that has attempted to provide only a theoretic treatment of the topic, we approach this question with microdata from deep learning, the AI paradigm responsible for nearly all landmark results in the past decade. We provide a framework for understanding the impact of two important trends: i) the recent breakthroughs using deep learning in R\&D, and ii) the rapid scaling of computation in deep learning systems. We show that if deep learning is widely adopted in the U.S. R\&D sector, it would induce an accumulation of computational capital that could nearly double the productivity growth rate.

Since the early 2010s, when it produced seminal breakthroughs in computer vision and speech recognition, deep learning has led to a rapid increase in the rate of progress in Artificial Intelligence (\cite{lecun2015deep}; \cite[Ch.\ 1]{goodfellow2016deep}; \cite[Ch.\ 1.3]{russell2002artificial}). Breakthroughs have been made in many areas, including, to name a few, computer vision, speech recognition, natural language processing, and game playing. Deep learning has also made inroads into parts of science largely untouched by previous AI research, including protein folding, semiconductor chip floorplanning, controlling nuclear fusion, and even discovering novel algorithms and new insights in pure mathematics. The rate at which long-standing problems have been solved, and the pace at which deep learning systems have out-competed traditional algorithms, have been surprisingly rapid to even some of its most seasoned practitioners. 

As uses of AI proliferate, economists have sought to understand its impacts on wages, factor shares, and economic growth. A prominent line of thought asks whether deep learning has the potential to become a General Purpose Technology, a technology with widespread applications in a variety of industries and the ability of AI to replace human labour across a wide variety of tasks (\cite{goldfarb2022gpt, agrawal_2022, trajtenberg_ai_2018}).\footnote{For example, \cite{goldfarb2022gpt} calculate the prevalence of deep learning-related job requirements in job postings within and across industries to predict whether technologies will become GPTs by whether they are in widespread use, capable of ongoing self-improvement, and enable multi-sector innovation. Across a set of 21 technologies, they find that machine learning displays the characteristics of a GPT in the deep learning era but not beforehand. In contrast, \cite{thompson2020computational} questions whether ongoing increases in deep learning performance will be sustainable, which could undermine its ability to provide the long-term benefits of a GPT.} 

Much of the existing research has focused on the potential of AI to impact final good production, but it has also been pointed out (e.g. by \cite{cockburn_4_2019}), that AI also has the potential to change the innovation process itself. Such ``Inventions of a Method of Invention'' (IMI) can significantly affect the rate of idea production (\cite{crafts_artificial_2021,cockburn_4_2019}) and, therefore, the overall rate of innovation in the economy. For example, building on the \cite{weitzman_recombinant_1998} model of recombinant technological development, \cite{agrawal_finding_2019} argue that deep learning can improve knowledge production by effectively searching through and recombining a wider range of ideas than is possible by human scientists and that this could result in accelerated economic growth. Empirical testing of the impact of AI on R\&D shows mixed results. \cite{bianchini_deep_2020} find that the use of deep learning is positively correlated with the mean and variance of paper citations received, increasing the likelihood for a contribution to become an influential ‘big hit.’ However, they also find that it is negatively correlated with the re-combinatorial novelty of ideas, measured as a function of the fraction of novel citation pairs in a given paper. Another line of research focuses on the relationship between AI and data, showing that machine learning increases the returns to data and thus the rate of knowledge production for data-rich firms (\cite{beraja_data-intensive_2020, abis_changing_2020}; \cite{agrawal2018prediction}). While these insights are informative about firm-level effects, they shed less light on the implications for the aggregate economy.

The impacts of AI on the innovation process deserve special attention because it has been pointed out that these, under suitable conditions, can have more dramatic permanent effects on productivity growth than those that arise from changes in final goods production. For example, in the semi-endogenous growth model of \cite{aghion_artificial_2019}, the authors consider AI automation in producing final goods and in producing knowledge, and find that the latter can produce much more rapid output growth. \cite{trammell_economic_2020} provide a review of the theoretical literature on AI and growth, which concludes that, while a high degree of automation in final goods production can produce a one-time increase in the growth rate, a high degree of automation in the R\&D sector can produce unbounded increases in economic growth.

Our work investigates how deep learning will impact the production of ideas. We argue that the adoption of deep learning makes computational capital in R\&D more productive, resulting in capital deepening that, if widespread, accelerates knowledge creation and economic growth. To motivate this, we derive a semi-endogenous growth model that shows that a positive shock to the R\&D elasticity of capital—such as might follow the widespread adoption of deep learning techniques—permanently increases the rate of idea accumulation and economic growth. 

But will deep learning increase the R\&D elasticity of capital? We provide supportive empirical evidence by estimating the idea production functions for two relatively mature deployments of deep learning. To analyze human capital in deep learning, we develop a novel machine learning approach for estimating human capital and apply it to machine learning papers in the arXiv repository. To analyze computing resources, we augment the dataset from \cite{thompson2020computational} to cover the entire universe of papers on two popular computer vision tasks.

Our estimates of the deep learning production function allow us to compare AI-augmented R\&D with the R\&D practiced in  U.S. science and engineering areas. We find that deep learning's idea production function depends notably more on capital. This greater dependence implies that more capital will be deployed per scientist in AI-augmented R\&D, boosting scientists' productivity and economy more broadly. Specifically our point estimates, when analysed in the context of a standard semi-endogenous growth model of the US economy, suggest that AI-augmented areas of R\&D would increase the rate of productivity growth by between 1.7- and 2-fold compared to the historical average rate observed over the past 70 years.

Our analysis is organized as follows. \hyperref[sec:Theory]{Section 2} motivates the importance of R\&D capital intensity for economic growth using a semi-endogenous growth model. \hyperref[sec:Data]{Section 3} describes the datasets, and \hyperref[sec:Empirical]{Section 4} the empirical strategy we use to model and estimate idea production. A key input for these models is an estimate of the human capital of the teams working on particular AI projects. In \hyperref[sec:Humancapital]{Section 5} we develop a deep neural network for learning simple representations of human capital that outperforms other measures commonly used in scientometric literature. For example, our human capital estimates explain 60-80\% of variance in key publication-related outcomes, whereas standard linear models explain less than 20\%. In \hyperref[sec:Analysis]{Section 6}, we present our empirical analysis, which implies that a firm in a competitive R\&D sector using deep learning would be roughly 5 times more capital-intensive than current U.S. STEM R\&D. In \hyperref[sec:Implications]{Section 7}, we use our growth theory model to investigate the implications of higher capital intensity of R\&D, and find that it implies a substantially faster productivity growth rate— 2- to 3-fold greater than the 0.8\% growth rate the US saw over the last decade. In \hyperref[sec:Robustness]{Section 8}, we find that our results are robust to outliers and alternative model specifications, including alternative elasticities of substitution between computational and human capital. In \hyperref[sec:Discussion]{Section 9}, we consider limitations of our analysis as well as future directions for research, and in \hyperref[sec:Conclusion]{Section 10} we conclude with a brief discussion of the implications of our results.

\section{Capital-intensive R\&D and deep learning}
\label{sec:capitalintenseRD}

\subsection{The role of capital in idea production}
\label{sec:capitalidea}

We argue that deep learning may affect the growth rate of knowledge by impacting the productivity of research capital. While the role of capital does not usually receive center-stage in the analysis of R\&D-based growth, it has been shown to generate permanent growth effects by increasing the marginal product of labour in R\&D and thus increasing investment in the R\&D sector (\cite{howitt_capital_1998, howitt_steady_1999}). The key mechanism driving this result is that, unlike the stock of human labour, the rate of physical capital accumulation can be readily increased or decreased in response to a change in its productivity. A very similar point is made by \cite{aghion_artificial_2019} in their study of the growth effect of AI. They show that in the classic endogenous growth case, a one-time increase in R\&D automation will raise the long-run growth rate, as capital—an accumulable factor in production—becomes permanently more important.

Empirical work has validated the importance of physical capital investment in idea production. For example, \cite{helmers2017my} showed that the creation of the UK's Diamond Light Source synchrotron increased the research capital available to local scientists, which in turn increased their research publication output relative to UK scientists located elsewhere.

The prevalence of capital goods in U.S. R\&D is documented by the National Science Foundation's Higher Education Research and Development Survey. It finds that academic institutions have spent over \$2bn per year on Capital Equipment or Software for R\&D since 2010, representing roughly 4\% of total direct R\&D expenditures (\cite{higher_survey}). In STEM fields, this share is higher—around 6\% overall in 2020, with Chemistry at 14\%, Material Science and Physics at 15\%, and Engineering ranging between 7\% and 11\%. Since the overall capital-intensity of academic science is only 4\%, there is enormous room for more capital-intensive approaches.

Even as compared to capital-intensive areas of science, there are reasons to suspect that R\&D using deep learning might be yet more so. Some recent capital-intensive examples include OpenAI's GPT-3 language model (\cite{brown2020language}), DeepMind's AlphaZero game-playing system (\cite{silver2017mastering}), and DeepMind's protein-folding system, each of which reportedly used millions of dollars worth of computing (\cite{gibney_self-taught_2017, jumper_highly_2021}).

Computing theory also suggests reasons deep learning is capital-intensive and why it is likely to become more so. In classical statistical learning theory, there generally is a trade-off between bias and variance (\cite{hastie2009elements}). Once a model grows beyond a certain complexity threshold, it tends to ``overfit" the data, worsening test performance\footnote{As distinct from the training error, test performance is calculated on data points never seen by the network.} as the variance term dominates. Deep neural networks seem capable of evading this trade-off by vastly expanding the size of the network (``overparameterization''), that is by deploying more computational capital (\cite{belkin2018reconciling, nakkiran2021deep}).  

Surprisingly, empirical analyses have shown that the performance gains that accrue to these networks with millions or billions of parameters are highly predictable. Generally, these analyses of ``neural scaling laws'' find that test error falls according to a power law in the scale of such models, and therefore in the amount of compute used (\cite{kaplan2020scaling, hoffmann2022training, hestness2017deep, sun2017revisiting, lepikhin2020gshard, li2020train, jones2021scaling, bahri2021explaining, sharma2020neural}).\footnote{See particularly, \cite{bahri2021explaining} for an account of the state of the existing work in their Related Works section.} Researchers and practitioners are harnessing these dependencies to get better performance.  For example, \cite{thompson2020computational} shows that progress is highly dependent on computational resources across a wide range of machine learning tasks. For image classification on the ImageNet database, $71\%$ of the variance in model performance is explained by the computation used. The importance of computing resources in deep learning was elegantly summarized by Rich Sutton (\cite{sutton_bitter_2019}), a prominent figure in the field of reinforcement learning, who wrote:

\begin{displayquote}
The biggest lesson that can be read from 70 years of AI research is that general methods that leverage computation are ultimately the most effective, and by a large margin... Seeking an improvement that makes a difference in the shorter term, researchers seek to leverage their human knowledge of the domain, but the only thing that matters in the long run is the leveraging of computation.
\end{displayquote}

If deep learning is indeed more capital-intensive, the investment dynamics implied by endogenous growth models would predict a rapid scale-up to have occurred in the computational capital being used in AI-based R\&D. \cite{sevilla2022compute} find exactly that: since the advent of deep learning, the growth in the amount of computational capital typically used in milestone models doubles roughly every 6 months, far outstripping the rate during previous eras of AI. So, while the size of capital investments made in deep learning systems are still small compared to, for example, those required for large-scale physics experiments, there are compelling reasons to believe that these models are capital-intensive and will continue to become more so.

\subsection{R\&D capital intensity in a semi-endogenous growth model}
\label{sec:Theory}

Consider a simple semi-endogenous growth model along the lines of \cite{jones_r_1995}. There are two sectors, a goods-producing sector where output is produced and an R\&D sector where additions to the stock of knowledge are made. A fraction $\alpha_l$ of the labour force is used in the R\&D sector and fraction $1-\alpha_l$ in the goods-producing sector. Similarly, fraction $\alpha_k$ of the capital stock is used in R\&D and the rest in goods production. We make similar simplifying assumptions as \cite{romer_endogenous_1990} by supposing that $\alpha_l$ and are $\alpha_k$ exogenous and constant for expositional clarity. Ideas are non-rivalrous and the full stock is used equally in both sectors of the economy. For further simplicity, we assume constant returns to scale in the production of final goods. The quantity of output produced at time $t$ is thus:
\begin{equation}
    \label{eq:1}
    Y(t) = \big[(1-\alpha_k) K(t)\big]^\alpha \big[A(t)(1-\alpha_l)L(t)\big]^{1-\alpha}, \hspace{0.15cm} \text{where} \hspace{0.15cm} \alpha \in (0,1),
\end{equation}
The production of new ideas depends on the quantities of capital and labour engaged in research and on the level of technology. We assume there are diminishing returns in the production of new ideas in inputs ($\beta + \theta <1$). This assumption ensures a unique steady-state growth rate, and prevents the growth rate from exploding as the R\&D inputs grow without bound. The idea stock grows as follows:
\begin{equation}
\label{eq:2}
\dot{A}(t) = B\big[\alpha_k K(t)\big]^{\beta}  \big[\alpha_l L(t)\big]^{\gamma} A(t)^\theta, \hspace{0.15cm} \text{where} \hspace{0.15cm} B >0, \beta,\hspace{0.15cm} \gamma, \geq 0 \hspace{0.15cm} \text{and} \hspace{0.15cm} \beta + \theta <1,
\end{equation}
where $B$ is a positive shift parameter. We further make the simplifying assumptions, not uncommon in the literature, that there is a constant saving rate $s$, and that capital depreciates at a constant rate $\delta$. Moreover, our model is a semi-endogenous one. Hence, we suppose that population grows at exogenous rate $n$. Thus capital and labour accumulation are described as follows:
\begin{equation}
\label{eq:3}
    \dot{K}(t) = s Y(t) - \delta K(t), \hspace{0.15cm} \text{and} \hspace{0.15cm} \dot{L}(t) = n L(t), \hspace{0.15cm} \text{where} \hspace{0.15cm} s,\delta \in (0,1) \hspace{0.15cm} \text{and}\hspace{0.15cm} n > 0.
\end{equation}
Using equations (\ref{eq:1}-\ref{eq:3}), we solve for the steady-state rates of growth in ideas and capital (denoted as $g^{*}_a$ and $g^{*}_k$ respectively.\footnote{A complete derivation may be found in \hyperref[sec:deriving_steady]{Appendix A1}.}) These are:
\begin{equation}
\label{eq:4}
    g^{*}_a =\frac{\beta + \gamma}{1-\beta -\theta}n, \hspace{0.15cm} \text{and} \hspace{0.15cm}  g^{*}_k = \frac{1-\theta + \gamma}{1-\beta -\theta}n.
\end{equation}

\noindent \textbf{Proposition 1.} Consider a shift in the technology of R\&D that creates a positive shock to the R\&D elasticity to capital and, at worst, a proportional negative shock to the R\&D elasticity to scientists. That is, consider a shift from an R\&D setting described by (\ref{eq:2}) to a setting described as follows:
\begin{equation*}
\dot{A}(t) = B\big[\alpha_k K(t)\big]^{\beta'}  \big[\alpha_l L(t)\big]^{\gamma'} A(t)^\theta
\end{equation*}
\begin{equation*}
\text{where} \hspace{0.15cm} \beta' > \beta, \hspace{0.15cm} \beta' - \beta \geq \gamma-\gamma', \hspace{0.15cm} \text{and} \hspace{0.15cm}  \beta' +\theta < 1.
\end{equation*}
Such a shift in the technology of R\&D has the following implications:
\begin{enumerate}[label=(\alph*)]
    \item the rate of idea accumulation is strictly and permanently increased, and
    \item the rate of economic growth is strictly and permanently increased.
\end{enumerate}

\noindent \textbf{Proof.} Define $\Delta_X \equiv X'-X$ as the difference between pre- and post-values of parameter $X$. By assumption, $\Delta_\beta \geq - \Delta_\gamma$. The steady-state rate of growth in ideas is strictly increased if:
\begin{equation}
\label{eq:5}
    \frac{\beta' + \gamma'}{1-\beta' -\theta}n  > \frac{\beta + \gamma}{1-\beta-\theta}n.
\end{equation}
which follows from the fact that $\Delta_\beta \geq - \Delta_\gamma$. This proves part (a) of the proposition. The proof of part (b) may be found in \hyperref[sec:AppendixA2]{Appendix A2}.\\

Proposition 1 states that if there is a shift to an R\&D setting with higher R\&D elasticity to capital, and if the R\&D elasticity to scientists is not disproportionately negatively affected, the steady-state rate of technological change and economic growth will strictly and permanently increase.

\newpage
\begin{figure}[h]
    \centering   
    \includegraphics[width=1\textwidth]{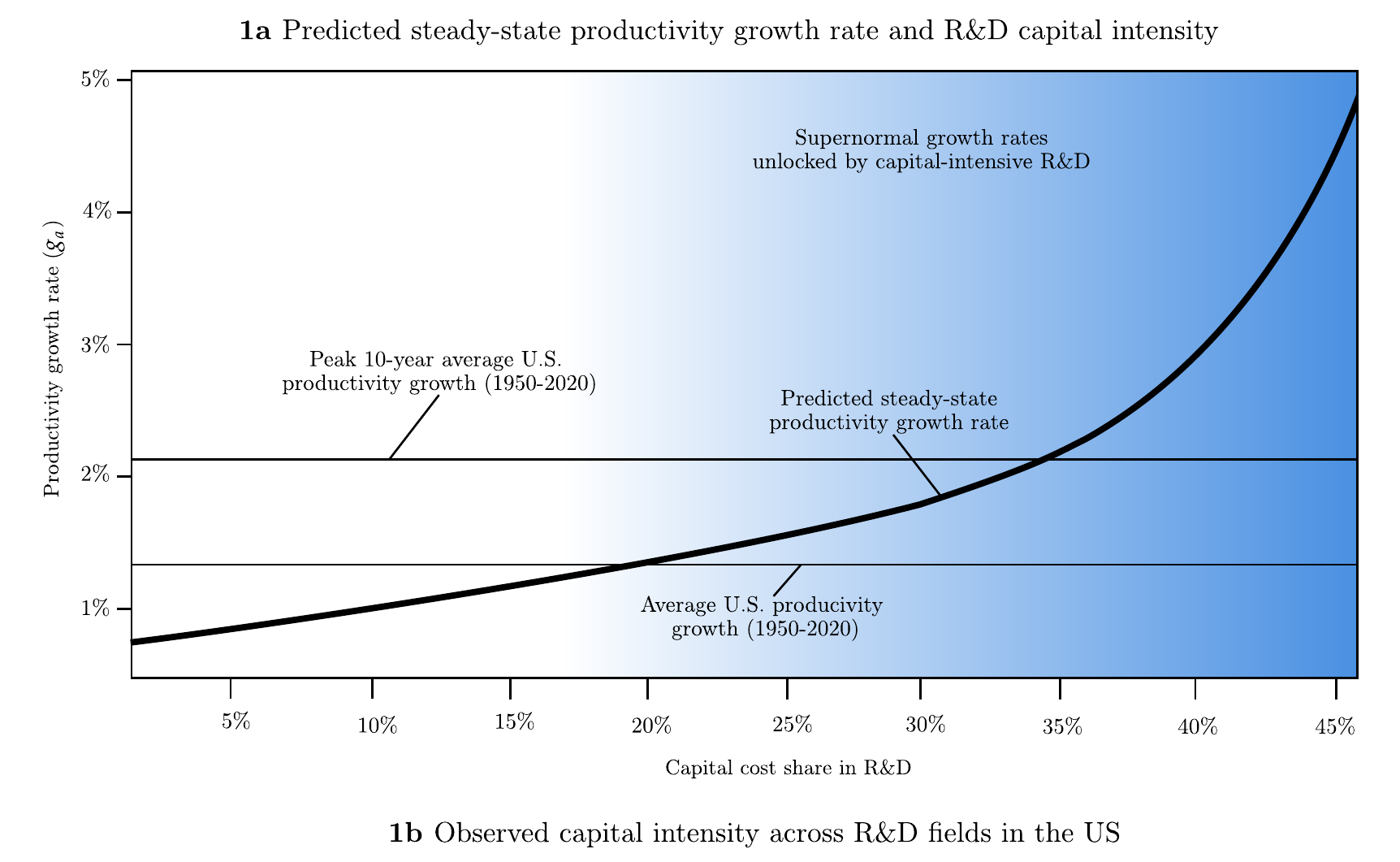}
    \vspace{-1cm}
    \label{fig:fig1a}
\end{figure}
\vspace{-0.3cm}
\begin{figure}[h]
    \centering   
    \includegraphics[width=1\textwidth]{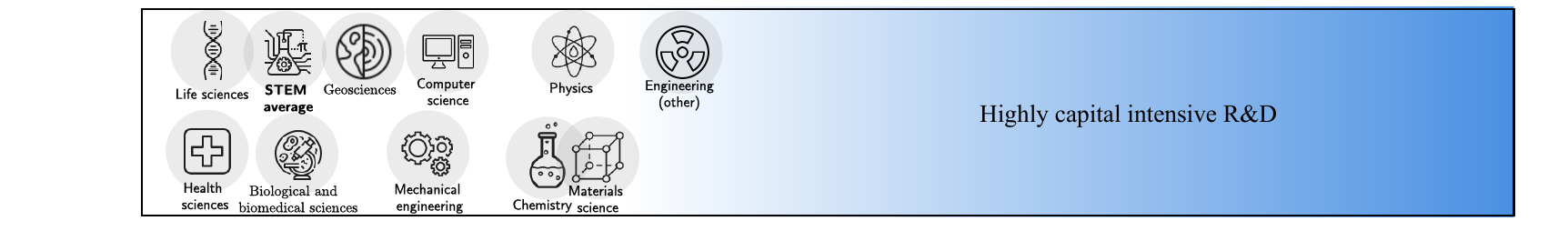}
    \vspace{-1cm}
    \label{fig:fig1a}
\end{figure}
\vspace{-.53cm}
\begin{figure}[h]
    \centering   
    \includegraphics[width=1\textwidth]{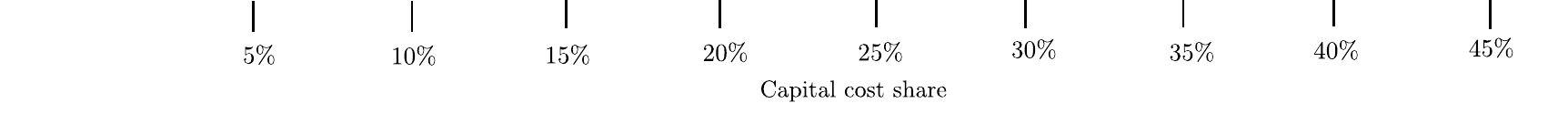}
    \vspace{-0.9cm}
    \caption*{\centering \small \textbf{Figure 1. Productivity growth and R\&D cost-shares in a competitive R\&D economy}. \textbf{Figure 1a} shows steady-state productivity growth as a function of the implied capital-cost share in a competitive R\&D industry when the elasticity of R\&D output to the stock of ideas ($\theta$) is $1/2$ and the elasticity of R\&D output to labour inputs ($\gamma$) is $2/5$ (consistent with our survey of existing estimates in \hyperref[sec:AppendixC]{Appendix C}). \textbf{Figure 1b} shows the share of R\&D expenditure in each discipline that is spent on capital equipment, based on data from the National Science Foundation 2020 Higher Education Research and Development Survey (\cite{higher_survey}).\footnotemark}
    \label{fig:fig1b}
\end{figure}
\footnotetext[5]{The underlying data and the details of the calculations used for Figure 1b may be found \href{https://docs.google.com/spreadsheets/d/1uno704Jd4vAFr1A_8M5XoVIXUKNsbbYBHLGB9l89Qf4/edit?usp=sharing}{here}.}

The intuition behind this result is as follows: while the labour force grows at a rate independent of economic growth, capital accumulation is determined endogenously by investment. After a positive shock to the R\&D elasticity to capital, investment rises, increasing the productivity of scientists. This increases the rate of idea production and consequently boosts economic growth. But faster economic growth also increases the rate of capital investment. Hence, a positive shock to the capital productivity of R\&D gives rise to a virtuous cycle of idea accumulation, economic growth and capital formation: a cycle that produces a new balanced growth path with permanently higher steady-state rates.

To illustrate how capital-intensive R\&D could result in super-normal productivity growth, consider \hyperref[fig:fig1]{Figure 1a}. Suppose the R\&D sector is competitive (such that wages and rents are equal to their marginal products), the steady-state rate of idea accumulation increases steadily in the share of R\&D expenditure dedicated to capital. Under conservative assumptions on our growth model, highly capital-intensive R\&D (such as when optimising R\&D firms dedicate at least 20\% of R\&D expenditure to capital) would produce productivity growth rates in excess of the usual productivity growth rates observed in the US. By contrast, current US R\&D tends to be highly labour-intensive. Using 2020 data from the \cite{higher_survey} NSF-supported STEM R\&D and assuming a Cobb-Douglas functional form for ideas production, we see that capital shares tend to fall between 3\% and 20\% (see \hyperref[fig:fig1]{Figure 1b}).\footnote{STEM fields tend to have \textit{higher} capital intensities than non-STEM fields.}\footnote{That is, to compute the capital share, we divide capital expenditures by total R\&D expenditures in each field. Our assumption and the resulting estimated cost shares are corroborated by \cite{czarnitzki_knowledge_2009}, who use a Cobb-Douglas specification and find that the capital share in R\&D among Belgian firms is less than 15\%.}

In the \hyperref[sec:Analysis]{analysis} section, we present evidence that the relative returns to capital for deep learning are higher than other types of R\&D. This suggests that, if deep learning could be similarly applied to a wide range of R\&D problems, its high degree of capital intensity could accelerate technological change and, as a consequence, economic growth.

\section{Data}
\label{sec:Data}

In our work, we rely primarily on two datasets. Our primary dataset covers the compute cost and performance for 151 deep learning models that were presented in publications between 2012 and 2021. The second is a bibliometric dataset of the authors of machine learning publications published between 1993 and 2021, which we use to infer the human capital inputs for each deep learning model in our primary dataset.

\subsection{Data on computer vision experiments}
\label{sec:DataCV}

Our dataset on the compute costs and performance covers 151 models published between 2012 and 2021. This data is an augmented version of \cite{thompson2020computational}, with additional details about the settings under which the models were trained and tested, for example, whether additional training data was used, or whether the training or test data was augmented.

The compute estimates are derived from the underlying papers following the procedure described by \cite{sevillacompute}, which we summarize in \hyperref[sec:AppendixE]{Appendix E}. The inclusion and exclusion criteria used to generate our datasets is described in \cite{thompson2020computational}. Deep learning models in this dataset span two well-known benchmarks: image classification on the ImageNet dataset and object detection on the Microsoft COCO dataset, usually known as the MS COCO dataset.

ImageNet is perhaps the most well-known and widely used computer vision dataset. It spans 1,000 object classes and contains 1.28m training images (\cite{ILSVRC15}). Some of the most important breakthroughs in deep learning have happened in ImageNet models, starting with AlexNet, a watershed moment when deep learning first outperformed other techniques on this task (\cite{krizhevsky_imagenet_2017}). Importantly for our purposes, success on ImageNet has often proven to be general: techniques that advance its state-of-the-art have usually been found to be successful in other tasks and domains. For example, \cite{beyer2020we} documents various instances when progress on ImageNet due to architecture design or optimization has yielded corresponding gains on other modalities, such as natural language processing, audio processing, and game playing. Because of this, it is plausible that our results for this benchmark could generalize to tasks and domains beyond computer vision. 

The MS COCO 2017 dataset is one of the most frequently used datasets for object detection, face detection, and pose estimation, among other tasks. It contains a total of 2.5 million labelled instances in 328k images (\cite{lin2014microsoft}). Like ImageNet, this dataset has been used as a test-bed of many influential innovations, such as \cite{he2016deep}'s Residual Network architecture, which has since become widely used in computer vision (see e.g. \cite{khan2020survey}).

While these two domains of computer vision are crucial test-beds for deep learning, it would be better if we considered a wider range of scientific and technical domains in which these techniques were applied. Unfortunately, this is difficult because of challenges for both inputs and outputs. For inputs, many deep learning papers fail to report even basic details of their computational usage. For outputs, some areas of deep learning struggle to define objective measures of performance. For example, how should one define the ``correct'' text summary of a picture?\footnote{See \cite{thompson2020computational} for a more in-depth discussion of these challenges.}

\subsection{Data on authors and publications}

Our dataset on machine learning publications comes from \href{https://arxiv.org/}{arXiv}, a pre-print server commonly used in various STEM fields, including computer science, and Scopus, Elsevier's abstract and citation database. Our dataset includes all papers on arXiv that were posted between 1993 and 2021 that are from the subfields typically associated with machine learning: Machine Learning (stat.ML), Artificial Intelligence (cs.AI), Computation and Language (cs.CL), Computer Vision and Pattern Recognition (cs.CV), and Learning (cs.LG). We match the authors of these papers to their corresponding entries in Scopus using a variety of string distance-based matching approaches.  This technique allows us to match 90.1\% of authors, and spot testing on 300 random matches shows that 96\% were correct.

With the connection between papers to their authors' publication histories, we construct a timeseries for each author that shows their number of publications, \textit{h}-index, and citations (excluding self-citations). We supplement this data with similar timeseries of grant funding for each author's institution and department over time from the \href{https://www.dimensions.ai/}{Dimensions} grant database, institutional rankings over time from \href{http://csmetrics.org/}{csmetrics.org}, and measures of the scientific influence of computer science journals and conferences using from \href{https://www.scimagojr.com/}{SCImago}. For full details on data collection procedures, see \hyperref[sec:AppendixD]{Appendix D}.

\section{Empirical strategy}
\label{sec:Empirical}
We assume, along the lines of the semi-endogenous growth model outlined above, that idea production using deep learning depends on three factors: labour (scientists' human capital), specialised capital goods (computational capital), and total factor productivity (the extant level of `ideas' or technology upon which researchers build):
\begin{equation}
\dot{A}(t) = B A(t)^\theta S(t)^\gamma C(t)^{\beta}, \hspace{0.15cm} \text{where} \hspace{0.15cm} t>0, \text{and} \hspace{0.15cm} X(0)>0  \hspace{0.15cm} \text{for any} \hspace{0.15cm} X \in \{A, S, C\}.
\end{equation}
where $\dot{A}(t)$ denotes the change in the stock of technology, $S(t)$ the total human-capital input of scientists, and $C(t)$ refers to the total capital inputs. To estimate this using data, we replace $\dot{A}(t)$ with a measure of the performance of deep learning models, $C(t)$ with data on computational inputs, and $S(t)$ with estimates of scientific human capital inputs.\footnote{This model, which is a Cobb-Douglas production function, implies a certain level of substitutability between computational capital and scientists. In \hyperref[sec:substitability]{Section 8.3}, we validate this model by showing that our estimates of the relevant elasticity of substitution support it.}

\subsection{Empirical specification}
Consider an economy where the level of technology grows exponentially on the balanced growth path in the way standardly assumed in growth theory models:
\begin{equation}
    A(t) = A(0)\mathrm{e}^{gt}, \hspace{0.15cm} A(0)>0.
\end{equation}
We do not observe technology directly. Instead, we observe performance on machine learning tasks. In these cases, the level of performance—usually measured as a type of predictive accuracy—falls on the unit interval. We assume that performance relates to technology according to the logistic function, reflecting that the most challenging parts of innovation are being able to make some initial headway with a problem and then perfecting it,
\begin{equation}
    P(t) = \frac{A(t)}{1+A(t)}.
\end{equation}
This is a similar assumption used to model how effort relates to various outcomes when the outcomes are bounded, such as in contests (\cite{vojnovic2015contest, baik1998difference}), conflict interactions (\cite{hirshleifer1989conflict, jia2012technologies}), and persuasion (\cite{skaperdas2012persuasion}). Beyond the fact that this is a relatively standard transformation that enables us to map progress in technology onto a bounded interval, there are two further motivating considerations.

Firstly, we show that this functional form implies a power-law between the scale of the compute deployed and the level of error achieved, which is in line with a robust finding of the relevant machine learning literature (e.g. \cite{hoffmann2022training}) (see \hyperref[sec:powerlaw]{Appendix A5}). Secondly, this functional form enables us to construct a simple empirical counterpart for technological progress, which we derive as follows. First, assuming that growth rates in adjacent periods are approximately equal (i.e. that $g_t \approx g_{t-1}$) it can be shown that proportional technological growth relates to performance improvements as follows (see \hyperref[sec:empiricalspec]{Appendix A4}):
\begin{equation}
    \label{eq:9}
    \frac{\dot{A}(t)}{A(t)} = \log\Bigg(\underbrace{\frac{P(t)}{P(t-1)}}_{\makebox[0pt]{\text{\footnotesize Proportional increase in accuracy}}}\Bigg) \hspace{0.8cm} + \hspace{0.8cm}  \log\Bigg(\underbrace{\frac{1-P(t-1)}{1-P(t)}}_{\makebox[0pt]{\text{\footnotesize Proportional reduction in error}}}\Bigg),
\end{equation}
which thus provides us an easy-to-interpret decomposition of technological progress in terms of the (logs of) the proportional reduction in error rate and the proportional increase in accuracy. 

Let $\tilde{g_t}$ denote the approximation of $g_t$ given $P(t)$, i.e. $\tilde{g_t} \equiv \log\bigg(\frac{P(t)}{P(t-1)}\bigg) + \log\bigg(\frac{1-P(t-1)}{1-P(t)}\bigg)$. We can write the empirical specification of our model as follows:
\begin{equation}
    \tilde{g_t} = A(t)^{\theta-1} S(t)^\gamma C(t)^{\beta}.
\end{equation}
We thus obtain an empirical specification of $\tilde{g_t}$ that we can ground in the relevant empirical data. When relating the model to data, time becomes discrete, and experiments are produced by research groups, which are indexed by $i \in \{1,...,N\}$.

Assuming a multiplicative error model, we specify the empirical counterpart of (10) as follows:
\begin{equation}
 \tilde{g}_{it} = A_t^{\theta-1} S_{it}^\gamma C_{it}^{\beta}\epsilon_{it}, \hspace{0.15cm} \text{where} \hspace{0.15cm} \log\epsilon_{it} \equiv u_{it} \sim N(0, \sigma^2).
\end{equation}
Taking logs of both sides, we have:
\begin{equation}
   \log \tilde{g}_{it} + = (\theta-1) \log A_t + \gamma \log S_{it} + \beta \log C_{it}+ u_{it}.
\end{equation}
We estimate the following model:
\begin{equation}
\label{eq:13}
  \begin{aligned}
 \log\underbrace{\tilde{g}_{it}}_{\makebox[0pt]{\text{\footnotesize Extent of tech. progress by $i$}}}  \hspace{0.75cm} = \hspace{0.75cm}  (\theta-1)\log\underbrace{A_t}_{\makebox[0pt]{\text{\footnotesize Extant level of technology}}} \hspace{1.1cm} + \hspace{1.1cm}
\gamma \log\underbrace{S_{it}}_{\makebox[0pt]{\text{\footnotesize $i$'s human capital}}} \hspace{1cm} +\\ \beta \log\underbrace{C_{it}}_{\makebox[0pt]{\text{\footnotesize $i$'s computational capital}}} \hspace{1.1cm} + \hspace{1.1cm}
\mathbf{\alpha} \underbrace{\mathbf{X}}_{\makebox[0pt]{\text{\footnotesize Vector of controls}}}  \hspace{1cm} +  \hspace{1cm} u_{it}.
  \end{aligned}
\end{equation}
which we can estimate in a pooled fashion with a time-fixed effect that captures $(\theta-1)A_t$ for $t \in \{1,...T\}$. By default, we will fix the time periods as years. In the robustness checks section, we show that shorter or longer time windows do not change our overall results.

A key basic that is evident from our estimation procedure (which we import from endogenous growth theory) is that knowledge is non-rivalrous. This assumption warrants some reflection. It assumes that researchers have access to a common stock of knowledge at any point in time and that the advancements made by researchers in one period are available to others in the next period. These assumptions seem broadly reasonable given the open research norms in machine learning (e.g. publishing on arXiv), providing access to code (e.g. via GitHub, \href{https://paperswithcode.com/}{Papers with Code}, etc.) as well as the common tools (e.g. PyTorch) that embed model implementations. And, indeed, companies that are notoriously closed-lipped about technology are nevertheless relatively open about AI research (\cite{ahmed2020democratization}).

If, nevertheless, there are groups that try to keep trade secrets, and those tend to be areas of particularly high human capital (as one might expect), then our model would implicitly treat such knowledge as an additional benefit of human capital. All else equal, this would make our results about capital-intensity underestimates of the true level. 

\subsection{Operationalizing innovations}
\label{sec:Operationalizing}

In estimating our model, we need to operationalize proportional performance improvements in terms of observables. We measure this using our baseline data, where authors of each paper have indicated the touchstone models in the literature whose ideas they are building on. That is, for any particular task, we define relative performance gains as follows:
\begin{equation}
    P_{t}/P(t-1) \equiv  P_{t}/\text{Baseline}_t.
\end{equation}
That is, $i$'s innovation is defined as the proportional improvement over the performance of a model that is considered, by the contemporaneous literature, to be a relevant baseline model. This operationalization is chosen for two reasons. Firstly, it is common practice to report these values in the machine learning literature, as the extent of innovations are often illustrated through comparisons to existing baseline levels of performance (\cite{armstrong2009relative, melis2017state, pressel2018baseline}). Secondly, this notion of an improvement over a model lines up well with the usual notion of the change in stock of knowledge in R\&D-based growth models, such as those from \cite{romer_endogenous_1990, grossman1994endogenous} and others; it represents the extent of the innovation of a new design relative to the existing stock of ideas. To find the appropriate baseline levels of performance, we survey the models that are used as baseline results in the relevant literature and take the median of their performance (See \hyperref[sec:AppendixD]{Appendix D1}).

\section{Modelling human capital}
\label{sec:Humancapital}

The final remaining piece needed to estimate the deep learning production functions for these computer vision tasks is to construct a measure of the scientific human capital used for each of these models. This measure should be predictive of outcomes that are strongly influenced by human capital and must be inferable from available data about the scientists' track records. In addition to overall predictiveness, it is particularly important that estimates are good for junior researchers, who contribute importantly to this young field.  

Prior work has used various measures of the quality or status of scientists and engineers, including impact-based metrics, such as citation counts (\cite{azoulay2019does, jones2014age, zucker2002labor}), the number of high-impact citations (\cite{azoulay2014matthew}) and bibliometric indices such as the \textit{h}-index (\cite{teplitskiy2019experts, fisman2018social, breschi2014inventor}). These approaches have substantial limitations as approaches to measuring the human capital of teams of scientists and engineers. As we shall see, almost all of these measures are only weakly predictive of key outcomes where we would expect scientific and technical human capital to be important, such as the number of citations the work will receive in the future, or the quality of the journal or conference the publication is to be published in. Moreover, impact-based metrics, such as citation counts or the \textit{h}-index generally assign low scores to junior researchers, as citations often take a long time to accrue following the publication of scholarly work.

Our strategy for modelling researcher human capital is as follows. We construct a deep neural network (DNN) and train it to develop a single-dimensional representation of the total quality-adjusted research input (``human capital'') that is highly predictive of key bibliometric and publication-related outcomes. Our approach implements an encoder that maps many features about the publication's authors input to a single-dimensional representation, and a decoder model (built explicitly to have the function as a linear regression) that maps this representation onto citation- and publication-related outcomes. Our approach exploits the ability of DNNs for nonlinear data compression of high-dimensional input features (see e.g. \cite{hinton2006reducing}; \cite{kramer1991nonlinear}).\footnote{Similar bottlenecks are used for a variety of feature-learning tasks, such as through under-complete auto-encoders (see e.g. \cite{bengio2013representation}).} Our approach finds human capital representations that are highly predictive of key bibliometric and publication-related outcomes, and that substantially outperform the typical approaches used in the literature.

\subsection{Our machine learning approach to estimating human capital}

\begin{figure}[h!]
    \centerline{\includegraphics[width=0.9\textwidth]{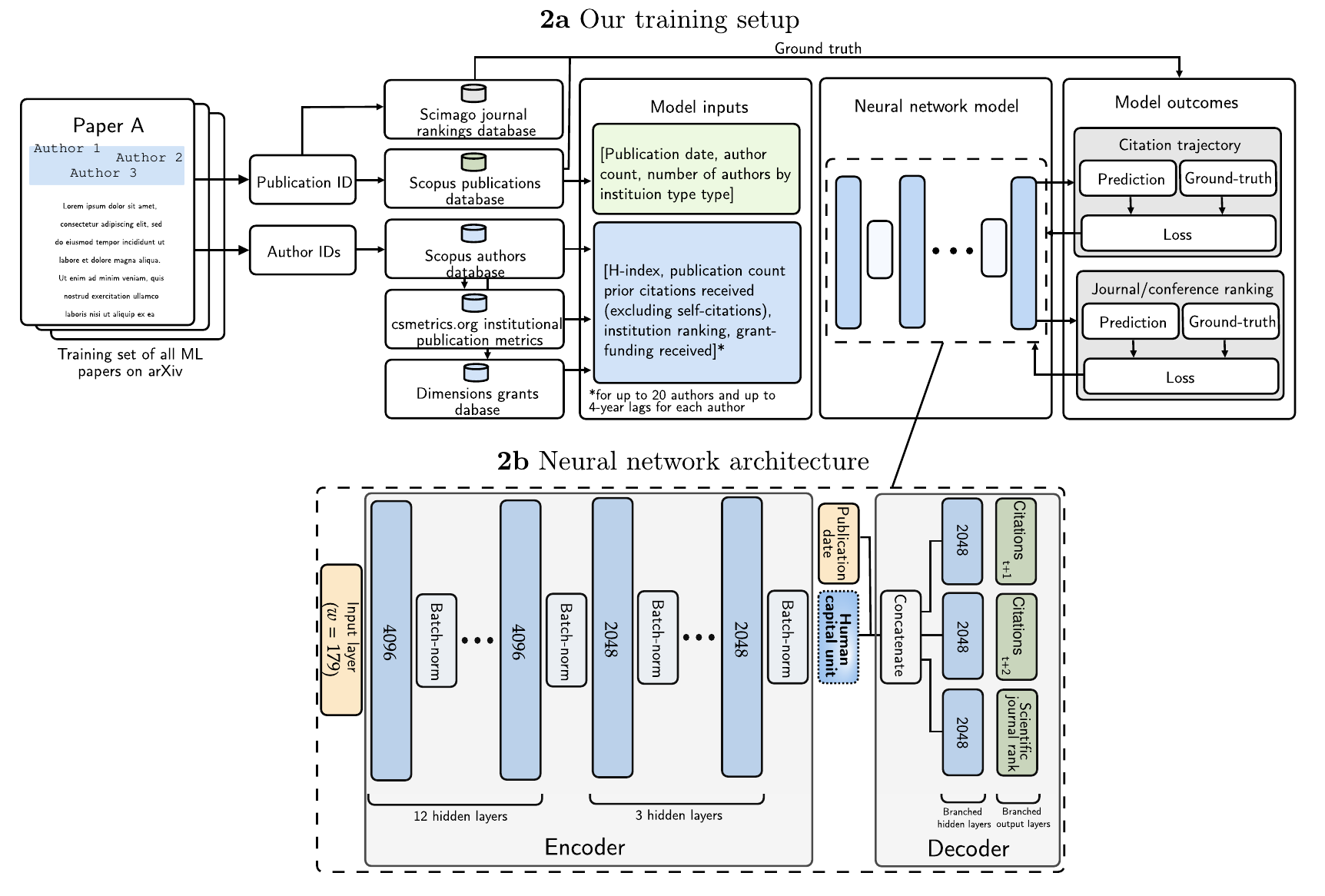}}
    \caption*{\centering \small \textbf{Figure 2. Human capital estimation strategy.}  \textbf{Figure 2a} presents our set-up for learning human capital representations for machine learning publications. \textbf{Figure 2b} shows our neural network architecture. Highlighted is the human capital unit, whose activations are strongly related to the quality of the research team. The numbers on each layer represent the number of units in that layer (for the human capital unit, this is just 1).}
    \label{fig:fig2}
\end{figure}

We use our dataset of 49,251 machine learning publications to train a neural network to predict bibliometric and publication-related outcomes. The predicted outcomes include the citation trajectories for each publication and its SJR-values, a measure of the quality of the journal or conference where the work ends up being published (see \cite{gonzalez2010new}). \hyperref[fig:fig2]{Figure 2a} provides a diagrammatic overview of our data pipeline and the training set-up used to produce our model. Further details of the training procedure may be found in \hyperref[sec:AppendixF]{Appendix F}.

Our architecture is constructed as follows. We first stack of 15 sets hidden layers, each consisting of a 4096 or 2048 node layer, followed by a batch-norm layer (\cite{ioffe2015batch}). These feed into a single unit — the ``human capital'' unit. This layer forces the neural network to reduce the dimensionality of its representations and distil the relevant features into a single scalar. The human capital result is then concatenated with the publication date, and fed into a series of independent sub-branches, one for each output being predicted. The final layer effectively implements separate linear regressions of the sort $y_i = \alpha + \mathbf{x}\beta$, meaning that the learned human-capital representations can only be linearly re-scaled and offset in order to make predictions about citations or journal quality. 

In other words, our approach implements an encoder that maps the input to the representation space, and a set of decoder regression models that map the representation onto citation- and publication-related outcomes. Thus, during training, the encoder is pushed to learn single-dimensional representations that are informative of human capital outcomes. 

\subsection{Validating our estimates}

\begin{figure}[h!]%
    \centering
    \subfloat[\centering Our human capital estimates predicts key outcomes much better than commonly used indicators]{{\includegraphics[width=10.7cm]{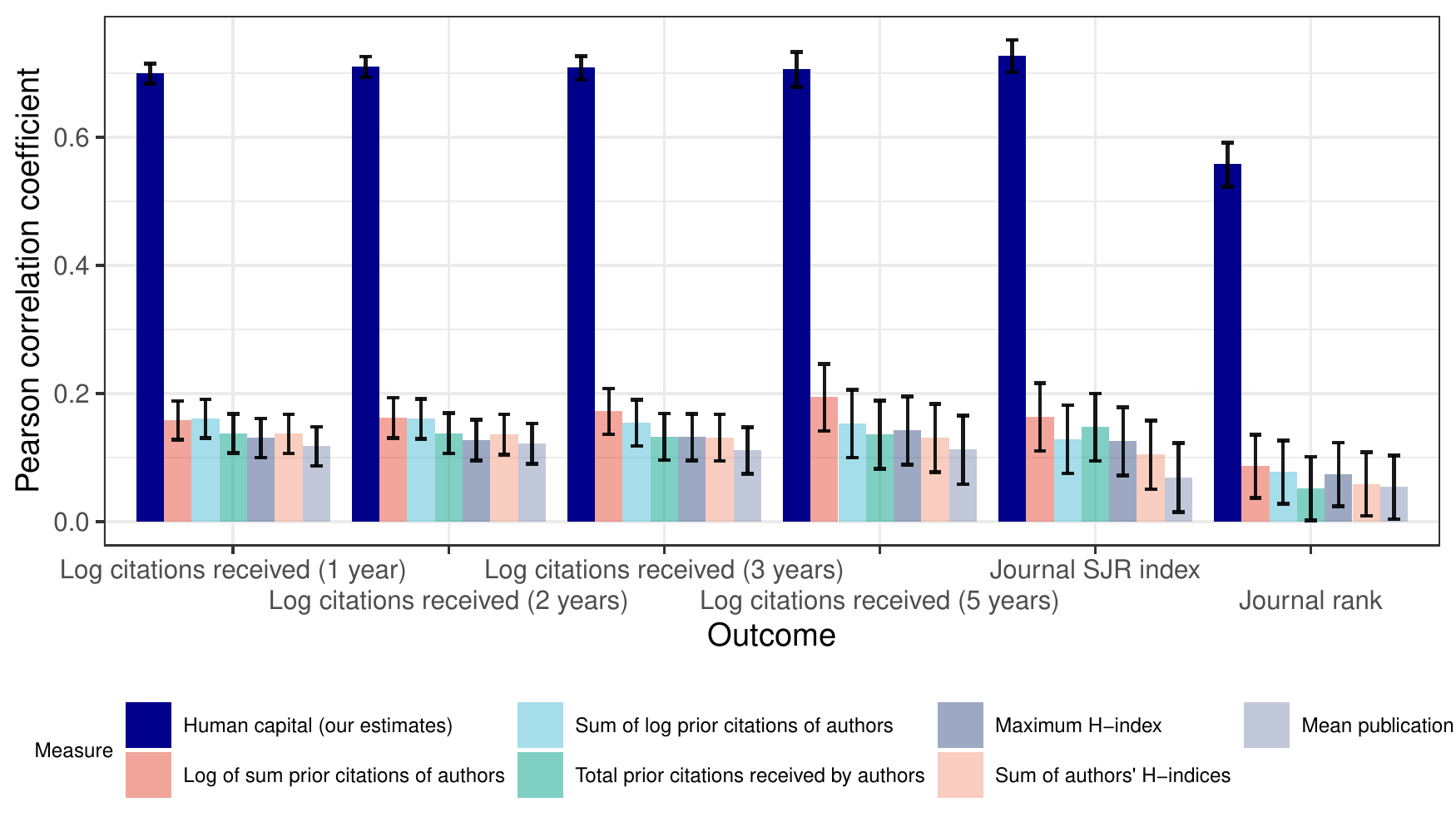} }}%
    \hspace{0.9em}
    \subfloat[\centering Our DNN achieves better accuracy compared to separate lasso regressions]{{\includegraphics[width=5cm]{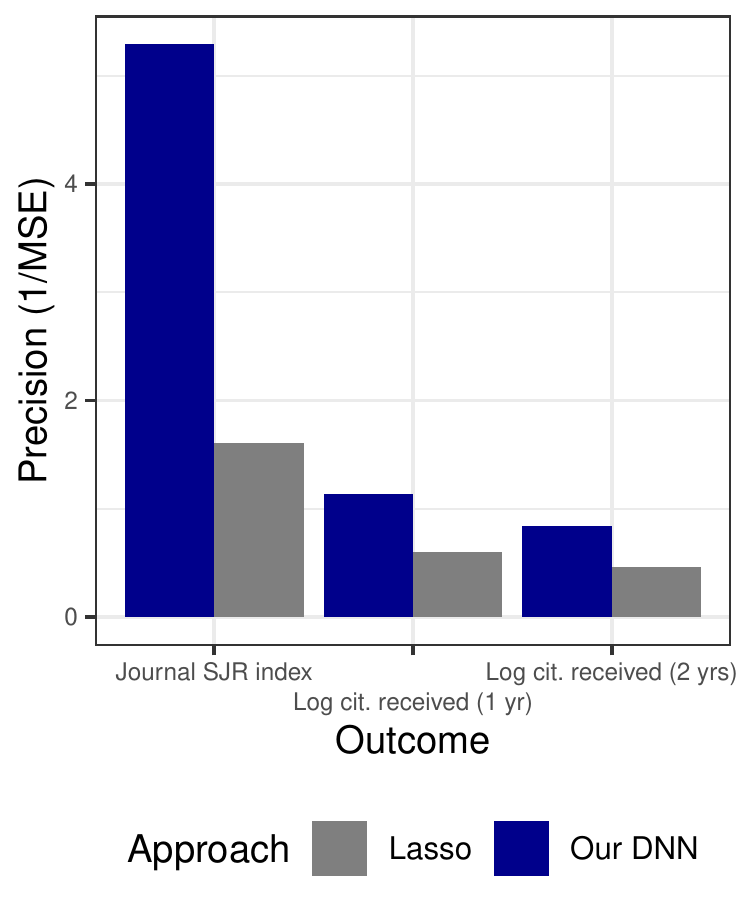} }}%
    \caption*{\centering \textbf{Figure 3. Evaluating our human capital measure}. \textbf{Figure (3a)}. Correlations between human capital measures and publication outcomes for a hold-out set of 4,081 publications. Error bars indicate the 95 confidence interval. Prior citations are the cumulative total citations received by the authors up until the year prior to publishing the relevant publication (excluding self-citations). H-indices and publication counts are evaluated at the year of publication for each author. Journal rank here represents the ordinal value of each journal in descending order of SJR-value. \textbf{Figure (3b)} shows the precision (defined as $1/\text{MSE}$) of predictions of our DNN-based model and Lasso regressions on the same hold-out set of 4,081 publications estimated separately for each outcome.}%
    \label{fig:fig3}%
\end{figure}

To assess the success of our measures in evaluating human capital, we compare their predictiveness across a range of outcomes, including citations received at various points and journal quality rankings.  In each case, our estimates predict more than 55\% of the variation in these measures, roughly 4-5 times as much as other proxies that are commonly used (e.g. prior publications, prior citations, \textit{h}-index) (see \hyperref[fig:fig3]{Figure 3}).  In all cases, we are predicting out of sample on a test set of a random sub-sample of 4,081 publications which was held out from any of the training.\footnote{Of these publications, incoming citations after 1 year are known for 4,035 publications, incoming citations after 2 years are known 3,724 publications, and the SJR values of the publication venues are known for 1,312 publications.}

These results indicate that our model has learned to predict bibliometric outcomes of publications, and in doing so, it has inferred meaningful and predictive human-capital features that can be measured as activation strength. Finally, when restricting the dataset to just publications with junior researchers (defined as publications with 2 or fewer prior publications), we find that our human capital estimates are still highly predictive of each of the bibliometric and publication-related outcomes (see \hyperref[sec:AppendixH]{Appendix H}), while most other proxies have little to no predictive power.

Thus far, we have shown that our measure has better predictive performance relative to other commonly-used human capital predictors. However, these other predictors also have access to much less data than our measure. For a more equal comparison, we also ask how our measure compares to a Lasso regression predictor—an approach more representative of linear approaches used in the literature—with access to the same inputs as our neural network. To do so, we evaluate our DNN on an out-of-sample test set. Our DNN represents a substantial improvement relative to simpler approaches found in the literature that rely on linear combinations of impact-based metrics, such as the \textit{h}-index, received publications, or publication counts. In particular, we obtain prediction errors (measured in mean-square-error) that are at least 40\% lower for each outcome compared to Lasso regressions, and thus we get much more precise predictions, as shown in Figure \href{fig:fig3}{3b}.

While the preceding points to clear benefits with our approach, it is essential to mention that our approach still has many of the limitations that are true for many measures in the field. First, there is no natural unit for human capital, and thus the cardinality of estimates is hard to interpret. Second, it is unclear how citations and journal quality relates to actual scientific merit, novelty, or insight. Thus, by using a measure predictive of citations and journal quality we would implicitly be missing aspects of human capital that cannot be inferred from these imperfect measures.

\section{Empirical Analysis}
\label{sec:Analysis}
Having validated our human capital measures, we combine them with the compute data to estimate production functions for two important AI tasks: image classification and object detection. In particular, we estimate individual regression models and a pooled model described by equation \ref{eq:13}. We estimate using OLS, except where a Breusch-Pagan test finds the presence of heteroskedascitity, in which case we estimate a GLS model by Maximum Likelihood (details in \hyperref[sec:AppendixJ]{Appendix J}).

\begin{table}[h!]
\centering
\small
\renewcommand{\arraystretch}{1.6}
\vspace{0.1cm}
\begin{tabular}{llcccc}
\toprule 
\multicolumn{1}{c}{Data} &
  Model &
  \multicolumn{3}{c}{Estimates} &
  \begin{tabular}[c]{@{}c@{}}Log\\ likelihood\end{tabular} \\ \hline
 &
   &
  \begin{tabular}[c ]{@{}c@{}} R\&D elasticity\\ to capital ($\beta$)\end{tabular} &
  \begin{tabular}[c]{@{}c@{}}R\&D elasticity to \\ human capital ($\gamma$)\end{tabular} &
  Trend &
 \\ \cline{3-6} 
\multirow{2}{*}{\begin{tabular}[c]{@{}l@{}} Image \\ classification\end{tabular}} &
  A1 &
  \begin{tabular}[c]{@{}c@{}}$\Underset{(0.021)}{0.111}$\signnn\end{tabular} &
  \begin{tabular}[c]{@{}c@{}}$\Underset{(0.086)}{0.246}$\signnn\end{tabular} &
  — &
  -48.798 \\
 &
  A2 &
  \begin{tabular}[c]{@{}c@{}}$\Underset{(0.029)}{0.140}$\signnn\end{tabular} &
  \begin{tabular}[c]{@{}c@{}}$\Underset{(0.085)}{0.350}$\signnn\end{tabular} &
  \begin{tabular}[c]{@{}c@{}}$\Underset{(0.014)}{0.051}$\signnn\end{tabular} &
  -39.787 \\ \renewcommand{\arraystretch}{2}
\multirow{2}{*}{\begin{tabular}[c]{@{}l@{}}Object \\ Detection\end{tabular}} &
  B1 &
  \begin{tabular}[c]{@{}c@{}}$\Underset{(0.106)}{0.246}$\sign\end{tabular} &
  \begin{tabular}[c]{@{}c@{}}$\Underset{(0.165)}{0.352}$\sign\end{tabular} &
  — &
  -43.256 \\
 &
  B2 &
  \begin{tabular}[c]{@{}c@{}}$\Underset{(0.090)}{0.253}$\signn\end{tabular} &
  \begin{tabular}[c]{@{}c@{}}$\Underset{(0.158)}{0.319}$\end{tabular} &
  \begin{tabular}[c]{@{}c@{}}$\Underset{(0.030)}{0.013}$\end{tabular} &
  -43.187 \\ \bottomrule
\end{tabular}
\caption*{\centering \textbf{Table 4. Deep learning production function estimates}. \small Estimation results for image classification ($n=96$) and object detection ($n=40$). \hspace{0.125cm}\sign,\hspace{0.125cm} \signn,\hspace{0.125cm} \signnn denote p$<$0.05, p$<$0.01, p$<$0.001 respectively.}
\label{Tab:table4}%
\end{table}
Our results for models A1-B2 are displayed in \href{Tab:table4}{Table 4}. We also estimate a pooled model with distinct time-fixed effects, which combines data across both computer vision tasks.  A likelihood ratio test indicates that the pooled model fits the data better than separately estimated models. The estimates of models C1-C2 are displayed in \href{Tab:table5}{Table 5}.  
\begin{table}[h!]
\centering
\small
\renewcommand{\arraystretch}{1.6}
\vspace{0.1cm}
\begin{tabular}{llcccc}
\hline
\multicolumn{1}{c}{Data} &
  \multicolumn{1}{c}{Model} &
  \multicolumn{3}{c}{Estimates} &
  \begin{tabular}[c]{@{}c@{}}Log\\ likelihood\end{tabular} \\ \hline
 &
  \multicolumn{1}{c}{} &
  \begin{tabular}[c]{@{}c@{}}R\&D elasticity to \\ capital ($\beta$)\end{tabular} &
  \begin{tabular}[c]{@{}c@{}}R\&D elasticity to \\ human capital ($\gamma$)\end{tabular} &
  Trend &
   \\ \cline{2-6} \renewcommand{\arraystretch}{1.6}
\multirow{2}{*}{\begin{tabular}[c]{@{}l@{}}Computer \\ Vision (pooled)\end{tabular}} &
  C1 &
  \begin{tabular}[c]{@{}c@{}}$\Underset{(0.022)}{0.145}$\signnn\end{tabular} &
  \begin{tabular}[c]{@{}c@{}}$\Underset{(0.088)}{0.278}$\signn\end{tabular} &
  — &
  -106.552 \\
 &
  C2 &
  \begin{tabular}[c]{@{}c@{}}$\Underset{(0.017)}{0.176}$\signnn\end{tabular} &
  \begin{tabular}[c]{@{}c@{}}$\Underset{(0.076)}{0.298}$\signnn\end{tabular} &
  \begin{tabular}[c]{@{}c@{}}$\Underset{(0.004)}{0.032}$\signnn\end{tabular} &
  -95.586 \\ \hline
\end{tabular}
\label{Tab:table5}
\caption*{\centering \textbf{Table 5. Pooled deep learning production function estimates}. \small Estimation results for pooled computer vision experiments ($n=136$). \signn,\hspace{0.125cm} \signnn denote p$<$0.05, p$<$0.01, p$<$0.001 respectively.}
\end{table}

For image classification, we estimate the R\&D elasticity of capital ($\beta$) is 0.111 (model A1) and 0.140 (model A2), as shown in \hyperref[Tab:table4]{Table 4}. This means that a 1\% increase in the computational capital used for this type of R\&D is associated with a 0.111-0.140\% increase in the rate of technological change. For object detection, we estimate $\beta$ is 0.246 (for both model B1 and B2), considerably higher than for image classification. For our pooled estimates, we get estimates between 0.145 and 0.176 (see \hyperref[Tab:table5]{Table 5}). All these results are statistically significant at the 5\% significance level; most are also significant at the 0.1\% level.

We find less variation in our estimates of the R\&D elasticity of human capital ($\gamma$) between the two deep learning tasks. Our human capital elasticity estimates for image classification are 0.246 (model A1) and 0.350 (model A2), both significant at the 0.1\% significance level. These estimates are just over twice as high as those for computational capital. For object detection tasks, the estimates of $\gamma$ are similar at 0.352 (model B1) and 0.319 (model B2), though only the former of these two estimates is significant at the 5\% significance level. For our pooled model, we find estimates of the R\&D elasticity of human capital is 0.278 (model C1) and 0.298 (model C2), each statistically significant at the 1\% significance level.

Recall that with the standard economic growth model outlined in \hyperref[sec:Theory]{Section 2.3}, we can directly infer the equilibrium cost shares dedicated to capital and labour from the relevant elasticities (assuming a competitive market). In doing so, we find that the implied capital-cost estimates range from 0.29 and 0.44 (see \hyperref[fig:implied]{Figure 4}). Our estimates indicate that an optimizing firm in the R\&D sector should allocate between 29\% and 44\% of their total expenditure on computational capital. Using our confidence intervals generated by bootstrapping, we find that implied capital-cost estimates of statistically significantly greater than 0.15 at the 5\% significance level for all models—a number that substantially exceeds most observed STEM R\&D capital shares.

\begin{figure}[h]
    \centering
    \includegraphics{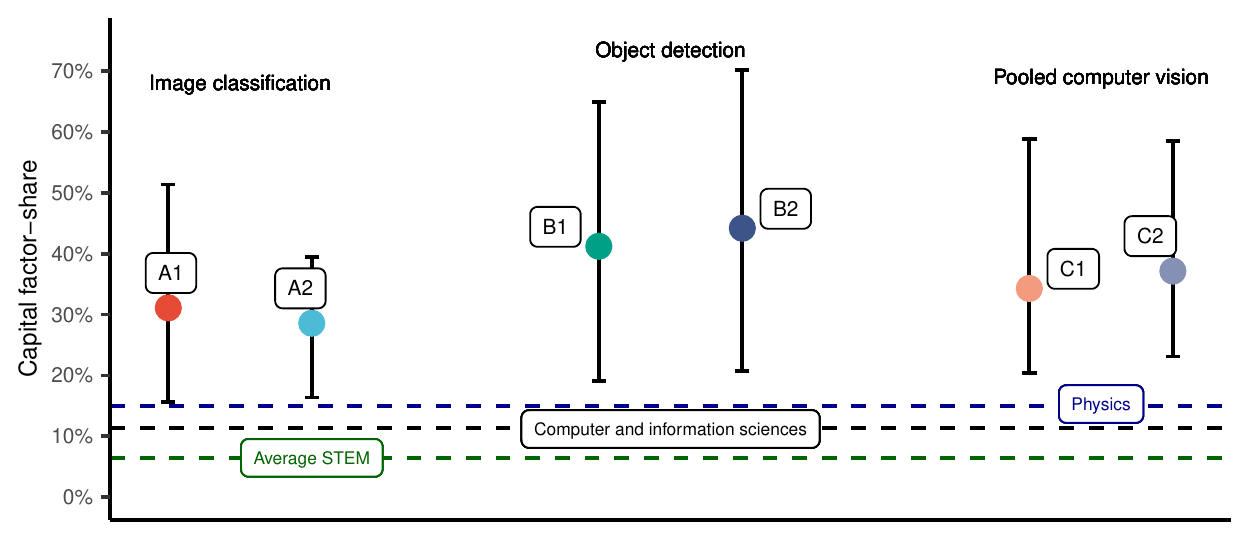}
    \caption*{\centering \small \textbf{Figure 4. Implied optimal R\&D expenditure breakdown}. Implied capital-cost shares given the estimates presented in tables \hyperref[Tab:table4]{4} and \hyperref[Tab:table5]{5}, computed as $\hat{\beta}/(\hat{\beta}+\hat{\gamma})$. Error bars represent 90\% confidence intervals generated by bootstrapping 10,000 iterations. We use the bias-corrected percentile method for bootstrapping confidence intervals for ratios outlined in \cite{campbell1999bootstrapping}.}
    \label{fig:fig4}%
\end{figure}

We find that the implied capital cost share for object detection is higher than for image classification by a margin of roughly ten percentage points. However, this difference is not statistically significant. Overall, we see that the implied R\&D capital shares for AI are substantially higher than in other areas of U.S. science and engineering \href{fig:fig1}{Section 2.3}, where capital share generally falls below 20\%.

\section{R\&D with deep learning}
\label{sec:Implications}

Having estimated the capital intensity of R\&D that is augmented with AI, we analyse the potential productivity effects that the widespread adoption of deep learning would have on economic productivity and growth. To do so, we suppose that the widespread adoption of deep learning would act as a one-time shock, raising the capital intensity of knowledge production in the economy to the levels we estimated in computer vision. 

Along the balanced growth path, the steady-state growth rate in the stock of knowledge is described by equation \ref{eq:4}. Using this, we can substitute in the parameter values implied by our empirical estimates from the prior section into our semi-endogenous growth model and compute the predicted change in R\&D productivity growth conditional on the widespread of adoption of deep learning. As shown in \href{fig:fig5}{Figure 5}, depending on the model specification, the results from image classification would imply a productivity growth rate between 1.6\% and 1.8\%, whereas those from object detection would imply a rate between 3.1\% and 3.9\%. Our preferred estimate, both because more data inform it and the estimates are more precise, is the pooled estimate for computer vision. With the widespread adoption of deep learning raising ideas production in the economy to this level of capital intensity, we would expect the productivity growth rate to rise to between 2.1\% and 2.4\%. To put that in context, this would amount to increase of between 1.7- and 2-fold relative to the 1.2\% average U.S. productivity growth from 1948 to 2021, and a 2.6- to 3-fold increase against the post-2000 0.8\% growth (\cite{goode_2012}). Thus, our results indicate that if adopting deep learning in other areas of R\&D allows those areas to leverage capital better in the same way that computer vision has, it will represent a substantial acceleration of scientific progress.

\newpage
\begin{figure}[h]
    \centering   
    \includegraphics[width=1\textwidth]{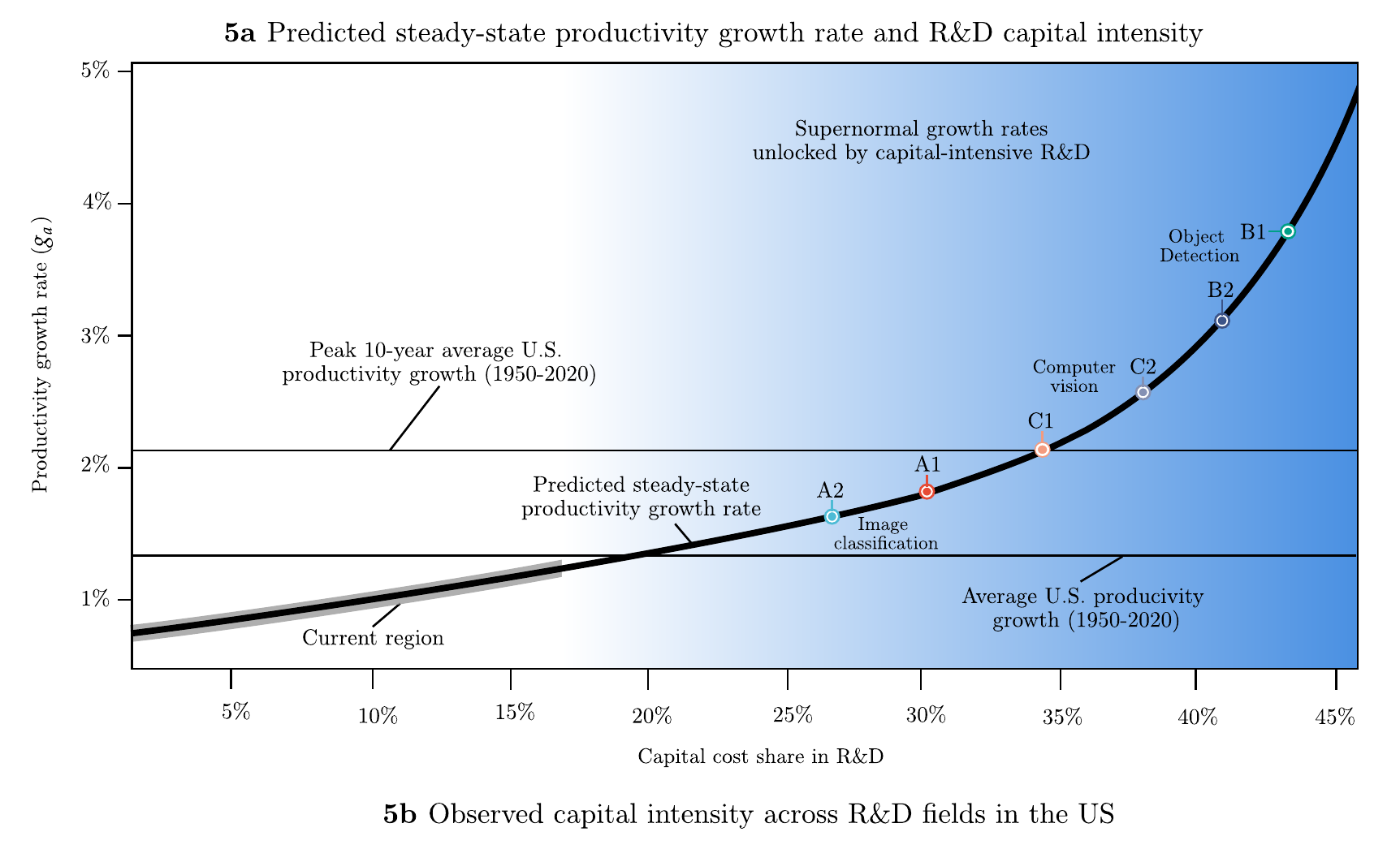}
    \vspace{-1cm}
    \label{fig:fig1a}
\end{figure}
\vspace{-0.2cm}
\begin{figure}[h]
    \centering   
    \includegraphics[width=1\textwidth]{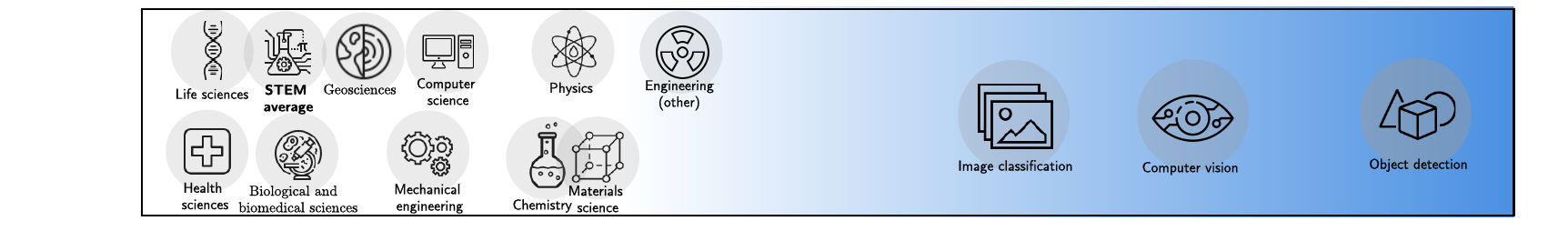}
    \vspace{-1.15cm}
    \label{fig:fig1b}
\end{figure}
\vspace{-.39cm}
\begin{figure}[h]
    \centering   
    \includegraphics[width=1\textwidth]{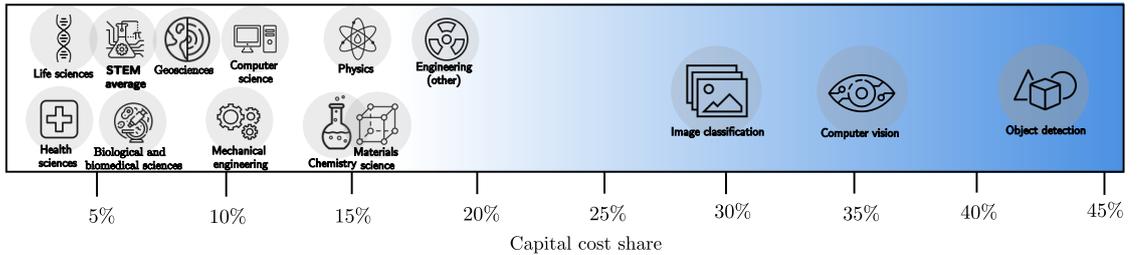}
    \vspace{-0.9cm}
    \caption*{\centering \small \textbf{Figure 5. Predicted productivity growth under widespread deployment of AI in R\&D}. Steady-state productivity growth as a function of the implied capital-cost share in a competitive R\&D industry when the elasticity of R\&D output to the stock of ideas ($\theta$) is $1/2$ and the elasticity of R\&D output to labour inputs ($\gamma$) is $2/5$. Markers indicate point estimates of implied optimal R\&D expenditure with deep learning according to models A1-C2 as estimated in section \ref{sec:Analysis}. ``Current region" indicates the current level of capital intensity of R\&D according to NSF data, which our semi-endogenous growth model predicts to result in 0.5\% to 1.3\% productivity growth, a level consistent with observed recent US productivity growth.}
    \label{fig:fig1c}
\end{figure}

\section{Robustness and external validity}
\label{sec:Robustness}

In this section, we show that the results in Section \ref{sec:Implications} are robust to outliers and to different choices of how granular or coarse-grained time periods are specified. Moreover, we show that our assumptions about the substitutability of labour and capital are consistent with our data, which supports a key assumption required for our inferences about the capital-intensity of deep learning from our estimated elasticities. Finally, we discuss the generalizability of estimates from computer vision to other R\&D tasks. 

\subsection{Sensitivity to outliers}

It is known, that certain empirical results from high-profile studies can be reversed by removing less than 1\% of the sample even when standard errors are small (\cite{broderick2020automatic}). In this section, we assess the sensitivity of our results to outliers by showing that the removal of a small fraction of the data is not determinative of our findings. To test the robustness of our results to the removing of samples, we re-run our analysis between around $10^4$ and $10^6$ times (depending on which model is re-estimated) on random sub-samples of our datasets that excludes a fraction of our observations. The point estimates are plotted in \href{fig:drop5percent}{Figure 6}.

\begin{figure}[h!]
    \centering
    \includegraphics[width=1\textwidth]{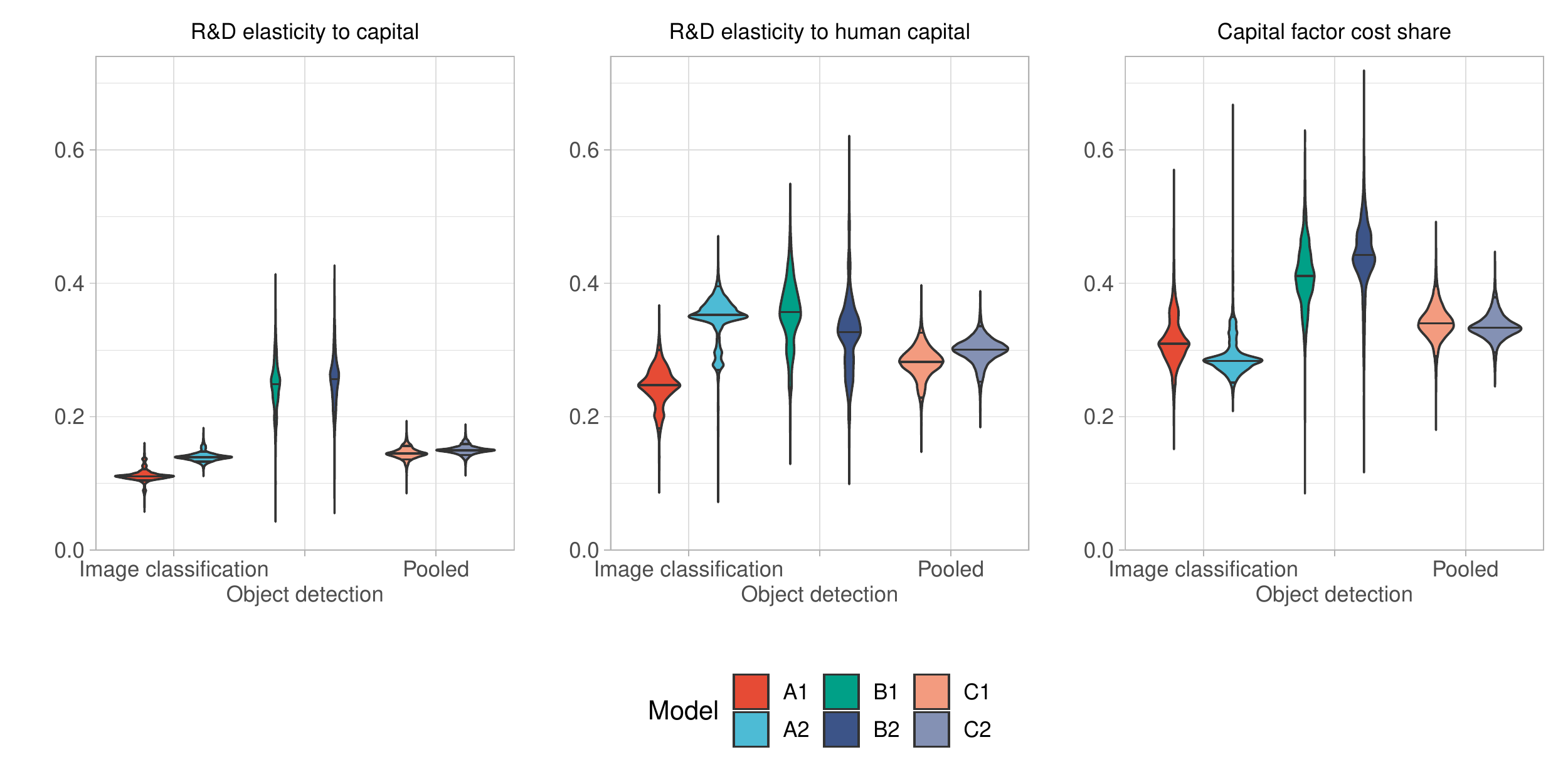}
    \label{fig:my_label}
    \vspace{-0.75cm}
    \caption*{\centering \small \textbf{Figure 6. Estimates after removing a random 5\% of dataset} Median estimates when taking a random sub-sample of our dataset that excludes 1.6\% to 5\% of the total number of observations. Results are displayed as violin plots, using kernel density estimation to create the distributions. Inside the violins, the box plots show median and interquartile ranges.}
    \label{fig:drop5percent}
\end{figure}
For our dataset on object detection, we estimated models on all possible sub-samples that exclude 3 observations (of which there are $\frac{40!}{37!3!}=9880$). For our dataset on image classification and for our pooled dataset, we estimated models on $10^6$ random sub-samples that exclude 5 observations. The sub-samples considered covers around 1.6\% of all possible sub-samples for the image classification dataset and 0.3\% of all sub-samples for the pooled dataset and is, therefore, a non-trivial fraction of all possible permutations.

We find that the point estimates are mostly robust to the removal of any small subset of observations, as we see that most estimates are tightly clustered around their median value, particularly for the estimates for which our datasets are the largest (namely, image classification and our pooled dataset). Moreover, the point estimates are consistent with the estimates found in our baseline empirical analysis presented in Section \ref{sec:Analysis}, indicating that our estimates are not the product of a small number of outliers.

\subsection{Alternative model specifications}
\label{sec:benchmark_windows}

Our estimation strategy for $A_t$ (the stock of knowledge) takes advantage of cross-sectional variation at each time point. That is, we effectively pool publications into groups of contemporaries published around the same time. We then estimate the variation in performance due to changes in inputs amongst these contemporaries—and suppose that this variation is due to inputs rather than changes in $A_t$. However, papers are published continuously over time, so this necessarily involves a bias-variance trade-off: if we specify more granular time periods (such as months instead of years), it gives more variance, but this will mean that each interval is estimated with fewer data points, making it noisier and more prone to over-fitting. In our empirical analysis, we balanced this trade-off by fixing time periods as yearly intervals. In what follows, we show that our conclusions are robust to different reasonable choices of how granular or coarse-grained periods are specified.
\begin{table}[h!]
\centering
\small
\renewcommand{\arraystretch}{1.6}
\begin{tabular}{lcccc}
\toprule
\multicolumn{1}{c}{Data} & \begin{tabular}[c]{@{}c@{}}Time-period\\ length\end{tabular} & Estimates & Estimates & \begin{tabular}[c]{@{}c@{}}Log\\ likelihood\end{tabular} \\ \hline
 &  & \begin{tabular}[c]{@{}c@{}}R\&D elasticity to \\ capital ($\beta$)\end{tabular} & \begin{tabular}[c]{@{}c@{}}R\&D elasticity to \\ human capital ($\gamma$)\end{tabular} &  \\ \cline{2-5} 
\multirow{3}{*}{\begin{tabular}[c]{@{}l@{}} Image \\ classification\end{tabular}} & 6 & \begin{tabular}[c]{@{}c@{}}$\Underset{(0.017)}{ 0.099}$\signnn\end{tabular} & \begin{tabular}[c]{@{}c@{}}$\Underset{(0.088)}{0.250}$\signn \end{tabular} & -47.379 \\
 & 12 & \begin{tabular}[c]{@{}c@{}}$\Underset{(0.021)}{0.111}$\signnn \end{tabular} & \begin{tabular}[c]{@{}c@{}}$\Underset{(0.086)}{0.246}$\sign\ \end{tabular} & -48.799 \\
 & 18 & \begin{tabular}[c]{@{}c@{}}$\Underset{(0.029)}{0.155}$\signnn \end{tabular} & \begin{tabular}[c]{@{}c@{}}$\Underset{(0.089)}{0.332}$\signnn\\ \end{tabular} & -75.380 \\
\multirow{3}{*}{\begin{tabular}[c]{@{}l@{}} Object \\ detection\end{tabular}} & 6 & \begin{tabular}[c]{@{}c@{}}$\Underset{(0.095)}{0.215}$\sign \end{tabular} & \begin{tabular}[c]{@{}c@{}}$\Underset{(0.139)}{0.400}$\signn\\ \end{tabular} & -43.256 \\
 & 12 & \begin{tabular}[c]{@{}c@{}}$\Underset{(0.096)}{0.246}$\sign \end{tabular} & \begin{tabular}[c]{@{}c@{}}$\Underset{(0.165)}{0.352}$\sign \end{tabular} & -43.256 \\
 & 18 & \begin{tabular}[c]{@{}c@{}} $\Underset{(0.111)}{0.214}$\nosign \end{tabular} & \begin{tabular}[c]{@{}c@{}}$\Underset{(0.204)}{0.348}$\nosign  \end{tabular} & -44.030 \\
\multirow{3}{*}{\begin{tabular}[c]{@{}l@{}} Computer \\ vision (pooled)\end{tabular}} & 6 & \begin{tabular}[c]{@{}c@{}}$\Underset{(0.016)}{0.150}$\signnn \end{tabular} & \begin{tabular}[c]{@{}c@{}}$\Underset{(0.072)}{0.294}$\signnn \end{tabular} & -74.276 \\
 & 12 & \begin{tabular}[c]{@{}c@{}}$\Underset{(0.020)}{0.132}$\signnn \end{tabular} & \begin{tabular}[c]{@{}c@{}}$\Underset{(0.075)}{0.245}$\signn \end{tabular} & -87.288 \\
 & 18 & \begin{tabular}[c]{@{}c@{}} $\Underset{(0.020)}{0.142}$\signnn \end{tabular} & \begin{tabular}[c]{@{}c@{}}$\Underset{(0.082)}{0.218}$\signn  \end{tabular} & -95.839 \\ \bottomrule
\end{tabular}
\caption*{\centering \small \textbf{Table 6. Estimation results for separate models with alternative window lengths}. Estimates of models A1, B1, and C1 with different window lengths. Specifications are the same as in the main analysis in Section} \ref{sec:Analysis}.
\label{tab:tab6}
\end{table}

We re-estimate models A1-C2 with window lengths 6, 12, and 18 months and compare our estimates to those obtained in Section \ref{sec:Analysis}. We find that the estimates are mostly similar for all relevant datasets (see Table \href{tab:tab6}{6} for estimates for models A1-C1, and appendix \hyperref[sec:AppendixK]{Appendix K} for re-estimates of models A2-C2). Moreover, similar patterns remain: estimates of the R\&D elasticity to capital ($\beta$) are relatively lower for image classification tasks than for object detection tasks. These estimates strongly suggest that our key estimates are robust to different choices of how granular or coarse-grained time periods are specified.

\subsection{Sensitivity to model assumptions about the substitutability of human scientists}
\label{sec:substitability}

In our semi-endogenous growth model, we assume that the elasticity of substitution equals 1 (i.e. is modelled by a Cobb-Douglas production function). Hence, we assume a substantial level of substitutability of human scientists for compute. In this section, we test: (i) whether our results would continue to hold if there was a lesser degree of substitutability of human scientists, and (ii) whether the data is consistent with a higher or lower level of substitutability.

The substitutability assumption is important because our semi-endogenous growth model implies that compute stock will grow faster than the stock of scientists. To see this, recall that from equation \ref{eq:4}, the steady-state growth of capital dedicated to R\&D is $\frac{1-\theta + \gamma}{1-\beta-\theta}n$ (which, with the estimates used in \href{sec:capitalintenseRD}{Section 2.2.}, this would be  $\sim 2n$, while the stock of scientists grows just at the rate $n$). Hence, we should expect that, in the fullness of time, the stock of specialized capital goods, $C(t)$, will be substantially greater than the stock of scientists $S(t)$.\footnote{This conclusion is consistent with what we see in our empirical data as well as what has been found in other areas of computing (\cite{thompson2022importance}).}

If human scientists were more difficult to substitute with compute than we have assumed, the optimal investment path might involve larger investments in human scientists.  We may ask: How much weaker of an assumption can we make for our conclusions to still follow, and is our assumption about the substitutability of human scientists reasonable? The first part is straightforward to answer. Suppose the idea production function instead followed a more general constant elasticity of substitution production function:
\begin{equation}
\label{eq:eq16}
    \dot{A}(t) = A(t)^\theta [\gamma S(t)^{\frac{\sigma-1}{\sigma}} + \beta C(t)^{\frac{\sigma-1}{\sigma}}]^{\frac{\sigma}{\sigma -1}},
\end{equation}
where $\sigma$ denotes the elasticity of substitution between compute and scientists. In this framework, assuming a competitive R\&D sector, the share of expenditure dedicated to compute (which we will denote by $f$) is given by:
\begin{equation}
    f = \frac{\beta C(t)^{\frac{\sigma-1}{\sigma}}}{\beta C(t)^{\frac{\sigma-1}{\sigma}} + \gamma S(t)^{\frac{\sigma-1}{\sigma}}}. 
\end{equation}
From this expression, we can see that in the typical Cobb-Douglas case (where $\sigma=1$), we have that $f=\beta/(\beta+\gamma)$. Note moreover, that whenever $\sigma\geq 1$ it must be that $f \geq \beta/(\beta+\gamma)$, while when $\sigma < 1$ it must be that $f < \beta/(\beta+\gamma)$. Hence, if scientists are less easily substitutable than we supposed, the capital intensity of R\&D will be lower than our estimates imply.

Hence, it is important to investigate whether $\sigma=1$ is a reasonable level of substitutability to assume. We investigate this by estimating equation \ref{eq:eq16}. Note that we can rewrite \ref{eq:eq16} by dividing by $A(t)$ as follows:
\begin{equation}
{g}_{t} = A_t^{\theta-1}[\gamma S_t^{\rho} + \beta C_t^{\rho}]^{\frac{1}{\rho}},
\end{equation}
where $\rho \equiv \frac{\sigma-1}{\sigma}$. To simplify the estimation procedure, we can approximate this expression using the second-order McLaurin expansion (i.e. the Taylor series evaluated at $\sigma=1$),
\begin{equation}
\log {g}_{t} \approx (\theta-1) \log A_t + \gamma \log S_t + \beta \log C_t + \frac{1}{2}\rho \gamma \beta[\log S_t - \log C_t]^2,
\end{equation}
This is simply the translog production function, which has the well-known advantage that it is linear in its parameters and, therefore, estimable using OLS. Because of this, this approximation is widely used in similar settings (\cite{guilkey_comparison_1983, berndt_translog_1973}). In our case, the empirical model we estimate becomes:
\begin{equation}
\log \tilde{g}_{it} = (\theta-1) \log A_t + \gamma \log S_{it} + \beta \log C_{it} + \frac{1}{2}\rho \gamma \beta[\log S_{it} - \log C_{it}]^2 + \epsilon_{it},
\label{eq:21}
\end{equation}
where each variable has its usual meanings it had in Section \ref{sec:Empirical}. 
\begin{figure}[h!]
    \centering
    \includegraphics[width=0.95\textwidth]{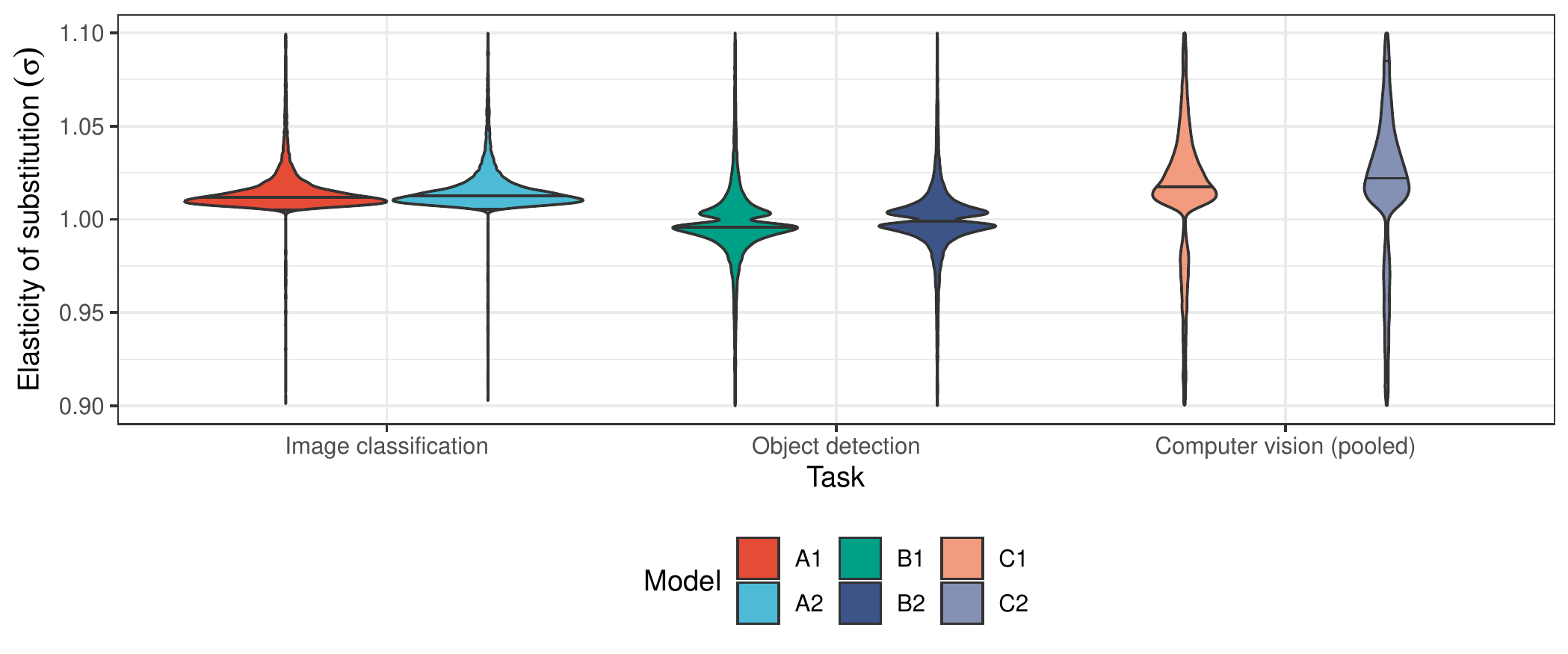}
    \label{fig:my_label}
    \vspace{-0.25cm}
    \caption*{\centering \small \textbf{Figure 7. Elasticity of substitution estimates} Kernel densities of the estimates of $\sigma$ across each of our main models generated by bootstrapping 10,000 iterations.}
    \label{fig:sigma}
\end{figure}
Figure \hyperref[fig:sigma]{6} plots the distributions generated by bootstrapping estimates of $\sigma$ in \ref{eq:21} for each of our models A1-C2. Our estimates of $\sigma$ are very tightly clustered around unity. These findings reinforce our assumption that $\sigma=1$ is reasonable and, therefore, our inferences about capital intensity are consistent with the data. Moreover, for the tasks for which we have the most data, we observe that our estimates $\sigma$ are slightly above $1$. Hence, for these tasks, it is likely that, if anything, our inferred level of capital intensity of deep learning-based R\&D is an underestimate of the true level after adequately accounting for the degree of substitutability of human scientists.

\subsection{Limits to external validity stemming from our choice of domain}
\label{sec:choicedomain}

Our analysis focuses exclusively on two computer vision tasks. Although this a small segment of scientific and engineering problems that deep learning might be applied to, there are good reasons to expect that the key insights gained from studying a wide range of architectures for computer vision to broadly could carry over to a wider range of R\&D problems.

One broad consideration in favour of this view is that deep learning often builds on common techniques, algorithms, and similar architectures across different subfields (see e.g. \cite[Ch.\ 3]{goodfellow2016deep}). Across many domains, deep learning systems are based on similar ideas and implemented using techniques and algorithms. At a high level, almost all modern deep learning systems—independently of the modality or task these are trained for—are all some `deep' computational graph with many parameters that are learned through gradient descent along gradients of some loss function computed by backpropagation. Indeed it is widely considered (e.g. by \cite{alom2018history}) that one of the first instances in which these key features of deep learning were all instantiated was with AlexNet in 2012, a model that is included in our dataset.

There certainly are some pronounced architectural divides across domains. For example, convolutional neural networks are widespread in computer vision, while transformer-based models are ubiquitous in machine translation. However, these architectural differences are also represented in our data. Indeed, our data includes convolutional neural networks, vision transformers, and architectures based purely on multi-layer perceptrons. Hence, much variation between modalities and tasks that exists within modern deep learning is reflected in the data that we consider. 

Some neural network architectures—such as the transformer, which is well-represented in our dataset—have been shown to operate effectively across domains, modalities and tasks (see, e.g. \cite{reed2022generalist}). This suggests that the upshot of our findings may well generalize to domains outside of computer vision. What is more, approaches like the transformer are often amongst the state-of-the-art techniques across many R\&D-adjacent tasks, such as code generation (\cite{li2022competition}), cheminformatics (\cite{irwin2022chemformer}), and bioinformatics (\cite{elnaggar2020prottrans}).

Moreover, work on neural scaling laws suggests that the relation between the model's size and performance scales according to the usual power-law independently of the domain the model is trained to handle. In fact, \cite{henighan2020scaling} find that for transformer models, there is a remarkable near-universal relation between the optimal model size and the compute budget across a range of domains spanning images, language, mathematics, video and more, which supports the notion the role of compute depends strongly on the type of technique or architecture, not the type of domain in which these techniques are applied. For these reasons, we expect that the key insights gained from studying a wide range of architectures for computer vision to broadly carry over to a wider range of R\&D problems to which deep learning is applied.

\subsection{How data availability influences our estimates}

One reason to expect that our results may fail to apply to a broader set of problems is that we studied a set of problems within computer vision with a relative abundance of quality labelled data. In some domains—such as the problem of protein folding (where the crystal structures of proteins are expensive and arduous to generate) high-quality data might be less readily available or expensive to generate. As a result of the relative abundance of data in computer vision, the returns to compute might be higher than they would be in lower-data-abundance regimes. To illustrate this, consider a simple model where deep learning system performance can be described as a constant elasticity production function in data $D$ and compute $C$:
\begin{equation}
\label{eq:20}
P = \big[\zeta D^{\frac{\sigma-1}{\sigma}} + \eta C^{\frac{\sigma-1}{\sigma}}\big]^{\frac{\sigma}{1-\sigma}}, \hspace{0.15cm} \text{where}  \hspace{0.15cm} \sigma,\hspace{0.05cm} \zeta,\hspace{0.05cm} \eta>0.
\end{equation}
It becomes clear whenever data and computation are gross complements ($\sigma<1$), then, the returns to compute will be lower in low-data regimes compared to high-data regimes:
\begin{equation}
    \frac{\partial P}{\partial C}\bigg\vert_{D_{\text{Low}}} = \eta C^{\frac{\sigma-1}{\sigma}-1} \big[\zeta D_{\text{Low}}^{\frac{\sigma-1}{\sigma}} + \eta C^{\frac{\sigma-1}{\sigma}}\big]^{\frac{\sigma}{1-\sigma}-1}\leq     \frac{\partial P}{\partial C}\bigg\vert_{D_{\text{High}}} = \eta C^{\frac{\sigma-1}{\sigma}-1} \big[\zeta D_{\text{High}}^{\frac{\sigma-1}{\sigma}} + \eta C^{\frac{\sigma-1}{\sigma}}\big]^{\frac{\sigma}{1-\sigma}-1}.
\end{equation}
This suggests that the returns to compute in high-data regimes can be unusually high. Moreover, empirical evidence has shown that AI-based ideas production is rapidly expanding in data-rich sectors such as investment management (\cite{abis_changing_2020}), and that computer vision firms with access to additional data are more innovative (\cite{beraja_data-intensive_2020}). Might our results, therefore, fail to generalize to low-data regimes? Our sense is that it might very well generalize, particularly for economically important R\&D tasks.

One piece of evidence for this view comes from work on neural scaling laws for language modelling. \cite{hoffmann2022training} derive a parametric loss function in which the amount of data and the model size enters additively separably. Although this result is derived in the context of language modelling, where data is abundant, this does suggest that data and compute can substitute for data by applying it to train larger models, at least when this training is done appropriately. 

Moreover, we might expect that for economically important R\&D tasks, complementary investments in generating the necessary datasets to train machine learning models will be made. For these tasks, we might expect such investments to produce high-fidelity physical simulations, high-quality synthetic datasets, the proliferation of sensors, and higher-throughput measurement apparatuses. Hence, insofar as we expect R\&D tasks to be of considerable economic importance, we might expect that low-data regimes are to be short-lived or otherwise atypical. For such tasks, a relative abundance of data may be representative.

\section{Discussion}
\label{sec:Discussion}

\subsection{Summary and implications}

Our main contributions are as follows. First, we provide a framework for understanding the impact of two important trends: i) the recent breakthroughs using deep learning in R\&D, and ii) the rapid scaling of computation in deep learning systems. We show that if deep learning is widely adopted in the R\&D sector and induces high returns to computational capital, then technological change will, under suitable conditions, permanently accelerate.

Secondly, using data from two computer vision tasks that are considered key test-beds for deep learning, we produce empirical estimates that imply that deep learning is more capital-intensive than other forms of R\&D; indeed an optimizing firm would dedicate between 29\% and 44\% of their total R\&D expenditure on (computing) capital. This result implies a clear empirical prediction: If deep learning is widely adopted in R\&D and this increases the returns to computational capital, then technological change will permanently accelerate. Consequentially, according to semi-endogenous growth theory, we will also get an acceleration of economic growth.

Thirdly, we make a methodological contribution by introducing a novel machine learning-based method for inferring human capital from scientific publications using an encoder to compress the inputs about the authors into a latent-space representation. In our approach, this encoder is tasked with learning representations of human capital that are most predictive of publication and citation-related outcomes that we have independent reasons to expect to be indicative of the quality of human capital. This human capital estimation predicts key outcomes 4-5 times more accurately than typical approaches in the literature.

Our work identifies three areas of future work that we expect to be fruitful. First, it would be valuable to better understand the diffusion of deep learning techniques within the economy, particularly within R\&D. We know that deep learning is diffusing to domains as diverse as protein-folding (\cite{jumper_highly_2021}), semiconductor chip design (\cite{mirhoseini_graph_2021}), and programming (\cite{li2022competition}). But how quickly this diffusion is happening and what drives it is poorly understood and micro-data charting it is lacking. Such analyses would inform us about the transition dynamics of AI-augmented R\&D across different industries. In particular, slower diffusion would imply a more protracted transition to the higher growth rates identified in Section \ref{sec:Implications}. Such analyses might also highlight areas where R\&D cannot be augmented with current AI techniques, implying more modest macro-level productivity improvements.

Another valuable area for research would be to look at whether improvements in AI techniques tend to make R\&D more capital-intensive. If true, the productivity gains could be more drastic. For instance, \cite{aghion_artificial_2019} shows that continually-increasing capital intensity in R\&D would result in ideas production becoming fully automated. So long as $\theta>0$ in Equation \hyperref[eq:2]{2}, this would produce unbounded growth. While we cannot analyze the possibility of this increase without further data or strong assumptions on our model, it seems worth studying whether the capital-intensity of deep learning varies with time.

A final area for follow-up work is how AI-augmented R\&D makes deeper changes to the knowledge production. While we have made progress investigating the relative importance of inputs to knowledge production and the implications this could have for the rate of technological change, many questions remain. In our work, we assumed that knowledge produced by labs using deep learning are essentially the same as that produced by unaugmented labs. Recall that we supposed that knowledge production was modelled as:
\begin{equation}
    \dot{A}(t) = A(t)^\theta S(t)^\gamma C(t)^{\beta},
\end{equation}
which implicitly assumes that $A(t)^\theta$—the intertemporal knowledge spillover for technological opportunities—is the same whether or not this knowledge was produced by AI-augmented R\&D or not. There might be reasons to be sceptical of this assumption in either direction. AI-augmented R\&D knowledge might be hard to share because deep learning is famously a `black box' technique that does not provide easy intuitions about what is being done. On the other hand, there is evidence that deep learning systems can learn essential building blocks (e.g. laws of nature (\cite{udrescu_ai_2020})) and that machine learning artefacts such as models are easy to disseminate widely (e.g. via Hugging Face (\cite{wolf_huggingfaces_2020})). These characteristics might make AI-augmented R\&D easier to diffuse, as has already been seen with the easy ability to adapt, fine-tune, or use existing models as `backbones' for adjacent problems. While it is unclear whether knowledge produced by AI-augmented R\&D has smaller or more significant spillovers, we expect it is worth investigating whether the extent of such spillovers depends on the technology used to generate knowledge.


\subsection{Selection issues and measurement error}

The data used in our empirical work may suffer from selection bias issues that are important to consider. Our dataset contains just models for which authors reported enough to infer how the model was trained and how much computation was needed. We might expect such reporting to be more common among computationally-intensive papers or papers whose performance is cutting-edge, which could introduce bias into our estimates—although it is not clear that this bias would be significant. If it did, it would likely positively bias our estimates of the returns to physical capital due to two effects. First, we suspect that publications that report these details are more likely to use large amounts of computation. Second, because the cost of deploying hardware for machine learning training is expensive (see e.g. \cite{sharir2020cost}), large compute deployments are likely associated with more effort to optimise training runs that ensures that these resources are used efficiently. Such efficiencies could lead to those deployments being better able to leverage computation to achieve particular results. As a consequence, our data would disproportionately be composed of those using computation more effectively. 

There are reasons to expect that such bias would be small. First, leading machine learning conferences require or at least encourage researchers to report details about hardware usage.\footnote{NeurIPS, which is one of the leading machine learning conferences, is an example of a conference that, as part of their mandatory checklist, currently includes questions about the hardware that is used (see the \href{https://neurips.cc/Conferences/2021/PaperInformation/PaperChecklist}{NeurIPS 2021 Paper Checklist Guidelines}). Requirements and norms such as these reduce the selection bias that stems from the fact that our dataset only contains models where the researchers reported compute-relevant information.} Second, the implementation of training runs are likely to be fairly similar across papers, as researchers typically use one of only a few types of open-source software that have distributed training settings and profilers that might be broadly comparable in achieving utilization. Moreover, we address this selection issue by using two different methods for inferring the amount of computation that was used in training (so that we could capture a wider range of models than would otherwise be possible). Hence, although selection bias might be introduced by the machine learning experiments included in our dataset, it is unclear whether this is a significant issue. Additionally, the variation in the amount of compute used spans over four orders of magnitude and will therefore dominate the variance in utilization rates, which suggests that this bias is likely to be small. Even if our estimate for the effectiveness of compute is overestimated, it is unclear whether this would imply that our estimate of capital share is over-estimated. This is because the expertise needed to optimize training runs will likely be captured in our human capital estimates so that these would also be overestimated, and it is unclear which effect would be more significant.

There are also potential measurement-error issues with our estimates of the amounts of training computation to train machine learning models. Firstly, we estimate only the amount of computation used in training the model in the final training run, not \textit{all} training runs. It is common for deep learning experiments to perform many (usually smaller) trial runs. These trial runs help inform the selection of hyper-parameters—parameters whose values control the learning process—used in the final training run. Our estimates of the compute that is used to train deep learning models are, therefore, likely to be a fraction of the total computation that was used. If this were a constant fraction, this would not bias our estimates of the relevant elasticities. It is currently not clear to the authors whether larger training runs use relatively more compute in selecting hyper-parameters than smaller training runs (in proportional terms). Regardless of the direction of this effect, the ratio of total computation used to the amount of computation used in the final training run will not be equal across experiments. This introduces noise in our estimates of elasticities, which will attenuate our estimates of the elasticity of computational capital downward.

Similar attenuation bias might be introduced in the process of estimating the computational inputs used to train deep learning models, as these rely on various approximations. To reduce the variance of our estimates' variance, we used multiple methods for producing these estimates and took the average of each. Given that the variation in the amount of compute used spans many orders of magnitude, minor errors due to the use of approximations are likely to be washed out. Nevertheless, our use of approximate techniques in estimations could have attenuated our estimates, which would mean that our results are underestimates.

\section{Conclusion}
\label{sec:Conclusion}

Much has been written on the potential mechanisms through which AI impacts productivity and output growth. Motivated by recent trends in deep learning, we provide empirical evidence for one of these mechanisms: the ability for scientists to efficiently harness more computational capital in R\&D. We argue that this mechanism is consequential because standard endogenous growth models imply that greater capital intensity in R\&D produces permanent increases in productivity growth and economic growth. We present evidence that deep learning-based computer vision research is significantly more capital-intensive than virtually all other R\&D sectors in the U.S.  If, as we argue, deep learning has similar impacts in other areas of R\&D, then the widespread adoption of deep learning-based R\&D could double U.S. productivity growth rates.

\printbibliography

\newpage

\section*{Appendix A: Derivations and mathematical results}
\label{sec:AppendixA}

\subsection*{A1: Deriving steady-state growth rates}
\label{sec:deriving_steady}

 Using (1-3), we can derive the rates at which capital and ideas grow:
\begin{equation}
    g_k(t) \equiv \frac{\dot{K}(t)}{K(t)} = c_k \bigg[\frac{A(t) L(t)}{K(t)}\bigg]^{1-\alpha} - \delta, \hspace{0.15cm} \text{where} \hspace{0.15cm} c_k \equiv s(1-\alpha_k)^\alpha (1-\alpha_l)^{1-\alpha}, 
\end{equation}
\begin{equation}
    g_a(t) \equiv \frac{\dot{A}(t)}{A(t)} = c_a K(t)^\beta L(t)^\gamma A(t)^{\theta-1}, \hspace{0.15cm} \text{where} \hspace{0.15cm}  c_a \equiv B \alpha_k^\beta \alpha_l^\gamma.
\end{equation}
Along the balanced growth path (defined as an equilibrium path where $Y(t)$, $K(t)$, $A(t)$ and $L(t)$ grow at a constant rate), it can be shown that:
\begin{equation}
   \tilde{g}_k(t) \equiv \frac{\dot{g}_k(t)}{g_k(t)} = (1-\alpha)(g_a + n - g_k),  \hspace{0.15cm} \text{and} \hspace{0.15cm} \tilde{g}_a(t) \equiv \frac{\dot{g}_a(t)}{g_a(t)} = \beta g_k + \gamma n - (1-\theta) g_a.
\end{equation}
The steady-state rates of growth in ideas and capital can then simply be found by solving for $g_k(t)$ and $g_a(t)$ that solves $\tilde{g}_k(t) = \tilde{g}_a(t) = 0$. Solving this system yields the following equilibrium growth rates (equilibrium growth rates are marked with the ${}^*$ superscript):
\begin{equation}
    g^{*}_a =\frac{\beta + \gamma}{1-\beta -\theta}n, \hspace{0.15cm} \text{and} \hspace{0.15cm}  g^{*}_k = \frac{1-\theta + \gamma}{1-\beta -\theta}n,
\end{equation}
which can be shown to be unique and stable.\footnote{This is a simple extension of the usual results for semi-endogenous growth models found in e.g. \cite[Chapter~3]{romer2012advanced}.}

\subsection*{A2: Proof of Proposition 1b}
\label{sec:AppendixA2}

For part (b) of Proposition 1, we wish to show that the steady-state rate of economic growth is also strictly increased under the relevant shift. This happens of both the steady-state rates of idea- and capital accumulation are permanently increased when $\Delta_\beta \geq - \Delta_\gamma$. Hence, to show this result, it suffices to show that the steady-state rate capital accumulation is permanently increased. This is true if:
\begin{equation}
    \frac{1-\theta + \gamma + \Delta_\gamma}{1-\beta' - \Delta_\beta -\theta}n > \frac{1-\theta + \gamma}{1-\beta -\theta}n
\end{equation}
Some rearranging reveals this to hold when:
\begin{equation}
    -\Delta_\gamma < \bigg(\frac{1-\theta+\gamma}{1-\beta-\theta}\bigg)\Delta_\beta
\end{equation}
Since $\frac{1-\theta+\gamma}{1-\beta-\theta} > 1$, this holds whenever $\Delta_\beta \geq - \Delta_\gamma > 0$. This yields the desired result. \qedsymbol{}

\subsection*{A3: Capital-cost share in competitive R\&D sector}

Given the competitiveness of the R\&D sector, total expenditure on wages and rents ($Lw$ and $Kr$ respectively) are given by:
\begin{equation}
    L w = L \frac{\partial \dot{A}(t)}{\partial L} = \gamma \dot{A}(t), \hspace{0.15cm} \text{and} \hspace{0.15cm} Kr = K \frac{\partial \dot{A}(t)}{\partial L} = \beta \dot{A}(t).
\end{equation}
From this it follows that the capital cost-share $Kr/(Kr + Lw)$ is given by $\beta/(\beta+\gamma)$.

\subsection*{A4: Empirical specification}
\label{sec:empiricalspec}
To see that 
\begin{equation}
    \frac{\dot{A}(t)}{A(t)} \approx \log \bigg( \frac{P(t)}{P(t-1)}\bigg) +\log \bigg(\frac{1-P(t-1)}{1-P(t)}\bigg).
\end{equation}
Note first that $\dot{A}(t)/A(t) = g_t$, and moreover that:
\begin{equation}
    \frac{P(t)}{1-P(t)}\frac{1-P(t-1)}{P(t-1)} = \frac{\exp(g_tt)}{\exp(g_{t-1}(t-1))} \approx \exp(g_t).
\end{equation}
Hence $\log \bigg( \frac{P(t)}{P(t-1)}\bigg) +\log \bigg(\frac{1-P(t-1)}{1-P(t)}\bigg) \approx g_t$ whenever $g_t \approx g_{t-1}$. 

\subsection*{A5: Our empirical model and scaling laws}
\label{sec:powerlaw}
One reason we assume the following relation between performance and technology, $P(t) = \frac{A(t)}{1+A(t)}$, is that this specification implies a power-law between the error-rate and the amount of compute dedicated to improving the underlying technology, which is a key finding found in the machine learning literature on `scaling laws' (such as \cite{kaplan2020scaling, hoffmann2022training}).\footnote{This connection was pointed out to us by Ege Erdil.} To show that this is the case, consider our usual idea production function, where, for convenience, we omit the input from scientists:
\begin{equation}
   \frac{\dot{A}(t)}{A(t)} = A(t)^{\theta-1}C(t)^\beta.
\end{equation}
\begin{equation}
\iff  \int \dot{A}(t)A(t)^{-\theta} dt = \int C(t)^\beta dt.
\end{equation}
We can solve this differential equation for $A(t)$,
\begin{equation}
 A(t) = \bigg[(1-\theta) \int C(t)^\beta dt + C\bigg]^{\frac{1}{1-\theta}}
\end{equation}
Taking the log of this expression yields:
\begin{equation}
 \log A(t) = \log\bigg((1-\theta) \int C(t)^\beta dt + C\bigg) - \log(1-\theta)
\end{equation}
\begin{equation}
\sim \log\bigg((1-\theta) \int C(t)^\beta dt \bigg)
\end{equation}
where $\sim$ means `is asymptotically proportional to'. Let $\sigma^{-1}$ denote the inverse sigmoid function, such that $\sigma^{-1}(x)=\log\big(\frac{x}{1-x}\big)$, then, noting that $\sigma^{-1}\big(P(t)\big)= \log A(t)$, we have:
\begin{equation}
P(t) \sim \sigma\bigg(\log\bigg((1-\theta) \int C(t)^\beta dt\bigg) \bigg).
\end{equation}
Using the fact that we can approximate the sigmoid function when the argument is large as $ \sigma(x) = \frac{1}{1 + \exp(-x)} \approx 1 - \exp(-x)$, we have:
\begin{equation}
P(t) \sim 1-\exp\bigg(-\log\bigg((1-\theta) \int C(t)^\beta dt \bigg) \bigg).
\end{equation}
\begin{equation}
\sim 1- \frac{1}{1-\theta}\int C(t)^{-\beta} dt.
\end{equation}
\begin{equation}
\iff E(t) \sim \frac{1}{1-\theta}\int C(t)^{-\beta} dt,
\end{equation}
where $E(t)$ denotes error and is defined as $1-P(t)$. Hence, consider a scaling of the compute path by some constant factor $\gamma$, then we find that
\begin{equation}
E(t) \sim \frac{1}{1-\theta}\int \big(\gamma C(t)\big)^{-\beta} dt = \frac{\gamma^{-\beta}}{1-\theta}\int C(t)^{-\beta} dt.
\end{equation}
That is, the `scale' of the path of compute dedicated improves the underlying technology (as measured in terms of the error rate that can be achieved) as a power-law. $\blacksquare$

\section*{Appendix B: Background on benchmarks in machine learning}
\label{sec:AppendixB}

Experimental benchmarks are a core feature of the way in which machine learning research is conducted. Benchmarks (a particular combination of a dataset or sets of datasets) have long been used for the purpose of performing experimental validation of new techniques (\cite{hothorn2005design}). It is currently not uncommon that the significance of new techniques are explicitly linked to benchmark performance, which often serves as the measuring stick for the amount of progress made (\cite{martinez2021research, raji2021ai}).

Benchmarks are a tool that enables the comparative assessment of competing machine learning techniques. For example, consider perhaps the most well-known computer vision benchmark, ImageNet (\cite{deng2009imagenet}). ImageNet\footnote{By ``ImageNet", we refer to ImageNet-1k, the  the most highly-used subset of the Scale Visual Recognition Challenge (ILSVRC) image classification and localization dataset, not the larger ImageNet-21K dataset.} is a set of over 1.2M images that belong to 1000 mutually exclusive classes. On the ImageNet Image Classification benchmark, deep learning models are trained to predict probability distributions over the classes of each image. Techniques are evaluated on a distinct test set consisting of 100k images, and typically evaluated on the basis of some top-$k$ error (the rate at which the `true' classes are not among the top $k$ highest-probability classes predicted by the model). A reduction in the error rate, then, represents some notion of progress at the level of this particular task.\footnote{That said, it should be noted that benchmark experiments, like those using ImageNet, have important limitations in assessing innovations for various reasons that are discussed in the machine learning Literature (e.g. \cite{raji2021ai, recht2019imagenet, picard2021torch}).}

\section*{Appendix C: Previous empirical analyses of idea production}
\label{sec:AppendixC}

In this section, we present a table that summarises our brief review of the literature that informed our choice of parameter values in our illustrations of the effects of an increase in the returns to capital on the rate of idea accumulation. 

\begin{table}[h]
\centering

\vspace{0.5em}
\small
\renewcommand{\arraystretch}{1.3}
\begin{tabular}{@{}lllll@{}}
\toprule
Study & Model & Location & $\hat{\gamma}$ & $\hat{\theta}$ \\ \midrule
\cite{abdih2006relating} &
\begin{tabular}[c]{@{}l@{}}``Restricted Cointegration\\ Model" (Table 3.)\end{tabular} & International & 0.21 & 1.44 \\
\cite{gong2004endogenous} & \begin{tabular}[c]{@{}l@{}}``Modified Romer model"\\ (Tables 1. and 2.)\end{tabular} & \begin{tabular}[c]{@{}l@{}} United States\\ and Germany\end{tabular}  & 0.10 to 0.48 &  0.1 to 0.83 \\
\cite{porter2000measuring} & \begin{tabular}[c]{@{}l@{}}``Direct Approach"\\ (Table 3.)\end{tabular} & \begin{tabular}[c]{@{}l@{}}OECD\\ member countries\end{tabular} & 0.39 to 0.48 & 0.84 to 1.19 \\
\cite{porter2000measuring} & \begin{tabular}[c]{@{}l@{}}``International spillovers and\\ R\&D productivity over time"\\ (Table 4.)\end{tabular} & \begin{tabular}[c]{@{}l@{}}OECD\\ member countries\end{tabular} & 0.30 to 0.85 & 0.02 to 0.90 \\
\cite{porter2000measuring} & \begin{tabular}[c]{@{}l@{}}``Regional, language and\\ trade-related spillovers" \\ (Table 5.)\end{tabular} & \begin{tabular}[c]{@{}l@{}}OECD\\ member countries\end{tabular} & 0.20 to 0.36 & 0.85 to 0.94 \\
\cite{kortum1993equilibrium} & \begin{tabular}[c]{@{}l@{}}``Estimates of the \\ Patent Equation" \\ (Table 4.)\end{tabular} & United States & 0.12 to 0.3 & — \\
\cite{kortum_inventions_nodate} & \begin{tabular}[c]{@{}l@{}} ``Estimates of the \\ Patent Equation" \\ (Table 4.1.)\end{tabular} & United States & 0.13 to 0.2 & — \\
\cite{hall_patents_1988} & \begin{tabular}[c]{@{}l@{}}``Estimates of the \\ Patent Equation" \\ (Table 5.)\end{tabular} & United States & 0.16 to 0.34 & — \\
\cite{hall_patents_1988} & \begin{tabular}[c]{@{}l@{}}``Estimates with Firm Effects" \\ (Table 6.)\end{tabular} & United States & 0.23 to 0.32 & — \\
\cite{hall_patents_1988} & \begin{tabular}[c]{@{}l@{}}``GMT Estimates \\ Assuming AR1 for R\&D" \\ (Table 7.)\end{tabular} & United States & 0.26 to 0.33 & — \\
\cite{pakes_patents_1984} & \begin{tabular}[c]{@{}l@{}}``Distributed Lag Estimates" \\ (Table 3.2.)\end{tabular} & United States & 0.56 & — \\
 \bottomrule
\end{tabular}
\caption*{\centering \small \textbf{Table 7. Previous empirical analyses of the idea production function}. Estimates of $\gamma$ and $\theta$ from our review of the relevant empirical literature. For each table, we provide the range of point estimates across models.} 

\end{table}

\section*{Appendix D: Data construction}
\label{sec:AppendixD}

In this section, we describe how each of our datasets were constructed. We provide a high-level overview of our methodology. For full details, see the \underline{\href{https://docs.google.com/document/d/1xZWALkldodSipA_qA1lsSaANrEo083QtD3nW4Z9BfqM/edit?usp=sharing}{online appendix}}.

\subsection*{D1: Data on deep learning models}

Our data of deep learning models was collated as follows. We use data on deep learning models for computer vision tasks from \cite{thompson2020computational}, which contains 46 publications with reported performance of the highest-performing model on the relevant benchmark, and estimates of the computational inputs required to train it. We supplemented this dataset with data on the test and training settings for each of these models, which we found by manually searching the publications. In case we could not find what training or test-settings were used, we dropped the model from out dataset.

We further collected publications from arXiv and Papers with Code that present models for either of two computer vision tasks anytime after 2012. For each of the best-performing models presented by each publication, we record the performance and calculate the computational inputs required to train the highest-performing model, using methods described in \cite{sevillacompute}. We further collected  the test and training settings for each of these models, which we found by manually searching the publications. In total, our dataset contains 148 models, their reported performance, estimates of the computational inputs required in training, and their test and training-settings.

\subsection*{D2: Data on baseline performance}
\label{sec:E2}

Our data on machine learning baseline results may be found \underline{\href{https://docs.google.com/spreadsheets/d/1dLP7T-A-Qo6Yl5OOdDC8B9dU41e5dGjiFYluy3J8C-g/edit?usp=sharing}{here}}. 

\subsubsection*{D3: Bibliometric panel dataset}

We generate a yearly panel dataset of feature for 223,703 authors on 115,235 machine learning papers from 1993 to 2021 using data from arXiv, a popular open-access repository for machine learning papers; Microsoft Project Academic Knowledge; and Scopus (\cite{noauthor_arxivorg_nodate, noauthor_project_nodate, noauthor_scopus_nodate}). Our panel dataset contains yearly series on all of the following author-specific features: number of publications, number of total citations, and authors’ \textit{h}-index (a measure of researcher productivity and impact, see \cite{hirsch2005index}). To build the dataset, we query arXiv for all papers from the following categories associated with machine learning: Machine Learning (stat.ML),
Artificial Intelligence (cs.AI), Computation and Language (cs.CL), Computer Vision and Pattern Recognition (cs.CV), and Learning (cs.LG). This produces  the entire universe of machine learning papers from 1993 to 2021. Using the procedure described in \cite{muller_is_2020}, we match these papers to their corresponding entries in the Microsoft Project Academic Knowledge database to obtain author names and affiliations. We match Scopus publications to those in the Microsoft dataset using a combination of DOIs and string-distance matching measures for their titles and author names. For each iterative step, we perform manual quality checks on a random sample of matches, the highest of which have false positive rates of 4\%. Through this procedure, all but 22,056 authors were matched.

To generate each series, we used bibliometrix, an R tool (\cite{aria2017bibliometrix}), to query Scopus’ API and retrieve all the publications of for each matched author in our dataset. For each publication, we then used pybliometrics, a Python interface to Scopus (\cite{rose2019pybliometrics}), to retrieve the citation trajectories—the number of incoming citations—for each publication per year, excluding self-citations. Authors' research fields were identified by retrieving Scopus’ subject area data on each of their publications. For each author, we further gathered data on fractional credit scores for citations and publications, which were generated by dividing the number of citations and publications by the number of authors on each publication.

\subsubsection*{D4: Grants dataset}

Data on grants received by institutions comes from the Dimensions database of academic publications (\cite{noauthor_dimensions_nodate}). We query the API for each institution in the Scopus panel dataset on author affiliations. This returns nominal data in USD on all grants per project-year and per institution-year. In order to make intertemporal comparisons, we deflate the institution-level data using the GDP implicit price deflator (\cite{us_bureau_of_economic_analysis_gross_1947}). We use the Damerau-Levenshtein measure of string distance to match the institution-year data to the matched Scopus-Microsoft cross-sectional paper titles dataset. For the top 1000 papers, we find 731 matches, of which 1.2\% are false positives. Overall, we match 50.8\% of institutions in our titles dataset. On a random sample of 100 matches, 5\% were found to be false positives. We report total grants received for the past 5 years by all authors’ departments or employers in real 2015 USD.

\subsubsection*{D5: Institutional rankings data}

Institutional rankings for Computer Science was generated using Computer Science publication data by csmetrics.org (\cite{noauthor_institutional_nodate}), an online dataset of institutional publication and citation metrics for computer science. This data is based on measured (retrospective) and predictive (prospective) metrics to compute a measure of publication impact of computer science publications by institution. Each institution in our dataset was matched to the entities in the csmetrics.org database by using Levenshtein distance, a measure of string-distance. Since not all institutions in our dataset were also present in csmetrics.org’s database, we were able to assign ranks to only 89.6\% of unique institutions in our dataset. The institutions we were unable to retrieve rankings for were smaller institutions that did not appear often in our data. In particular, institutional rankings were assigned in 99.4\% of authors for whom institutional data was available. Moreover, the matching rarely produced false-matches: of a random sample of 120 matches of institutions, only 0.9\% of matches were found to be incorrect.

\subsubsection*{D6: Computer science publication venue rankings data}

Data on yearly rankings of Computer Science publication venues was retrieved from SCImago Journal \& Country Rank (\cite{noauthor_scimago_nodate}), a portal for scientific indicators developed from the information contained in the Scopus database. In particular, we use their SJR indicator, which ranks scholarly journals, conferences and proceedings based on citation weighting schemes and eigenvector centrality, to quantify the impact and prestige of publication venues (\cite{gonzalez2010new}). Their database contains 5147 unique Computer Science venues spanning the 1999-2020 period. We used the Levenshtein distance to match the journal in SCImago’s database to the journals in which the articles dataset were published in. Overall, we were able to assign roughly 15,000 publications with journal rankings and SJR indicators. In case rankings were missing for the exact year of publication, we assigned each journal the rank of either its rank in the year prior to publication, or its rank in the year post-publication. If asynchronous rankings one-year apart were not available, we considered two-years prior and post-publication. Only 0.6\% of ranking assignments were made with asynchronous rankings of the kind described. Of a random sample of 100 publications, we found no incorrect matches between the Scopus-reported publication venue and the entity in SCImago’s database to which it was matched, suggesting our matching was performed highly accurately.

\section*{Appendix E: Estimating compute}
\label{sec:AppendixE}

We use two methods for inferring the amount of compute used to train an AI system, both are from \cite{sevillacompute}. These methods (one based on the architecture of the network and number of training batches processed; and another based on the hardware setup and amount of training time), are outlined below.

\subsubsection*{E1: Method 1—counting operations in the model}

The first method is can be summarized as:
\begin{equation}
    \text{Training compute} = (\text{FLOP per forward pass} + \text{FLOP per backward pass}) \times \text{Nr. of passes},
\end{equation}
where $\text{FLOP per forward pass}$ is the number of floating point operations in a forward pass, $\text{FLOP per backward pass}$ is the number of operations in a backward pass and $\text{Nr. of passes}$ is the number of full passes (a full pass includes both the backward and forward pass) made during training. 

Moreover, we use the facts that:
\begin{equation}
 \text{Nr. of passes} =\text{Nr. of epochs} \times \text{Nr. of examples},
\end{equation}
\begin{equation}
\text{FLOP per forward pass} + \text{FLOP per backward pass} \approx 3 \times \text{FLOP per forward pass}
\end{equation}
where the latter is implied by the fact that computing the backward pass requires each layer to compute the gradient with respect to the weights and the error gradient of each neuron with respect to the layer input to backpropagate. Each of these operations requires compute roughly equal to the amount of operations in the forward pass of the layer. Hence, the total number of FLOP per forward and backward pass is roughly 3-fold the number of FLOP per forward pass.

If the number of examples is not directly stated, it can be computed as the number of batches per epoch times the size of each batch: $\text{Nr. of examples} = \text{Nr. of batches} \times \text{batch size}$.

\subsubsection*{E2: Method 2—GPU time}

Secondly, we may also use the reported training time and GPU model performance to estimate the training compute. For example, if the training lasted 2 days and a total of 5 GPUs were used, that equals 10 GPU-days. By multiplying the number of GPU-days by the performance of the GPU, we can infer the amount of compute, in FLOP, that was needed to train the model. In particular, the formula used is the following:
\begin{equation}
    \text{Training compute} = \text{Training time (in seconds)} \times \text{Nr. of cores} \times \text{Peak FLOP/s} \times \text{utilization rate},
\end{equation}
where peak performance in FLOP/s is found in the relevant GPU product datasheets, and the utilization rate corrects for imperfect hardware utilization, for which 30\% is often a reasonable baseline (see \cite{sevillacompute}).

\section*{Appendix F: Training procedure and hyper-parameter settings}
\label{sec:AppendixF}

The weights in the first 12 layers are frozen, and the weights in the remaining components are randomly initialized and trained on a subset of for the three output features are known ($\sim$13k publications). Finally, we transfer this model to our dataset of publications that present machine learning models (which were removed from our training set), and inspect the activations of the human capital unit.

To train our DNN, we use the adamax optimizer (\cite{kingma2014adam}) with $\beta_1 = 0.9$, $\beta_2 = 0.999$. Throughout, use the GELU activation function (\cite{hendrycks2016gaussian}), with the exception of the Human Capital unit, which is ReLU-activated.
\begin{table}[h!]
\centering
\small
\renewcommand{\arraystretch}{1.6}
\begin{tabular}{@{}cccccc@{}}
\toprule
\multicolumn{1}{l}{Pre-training step} & \multicolumn{1}{l}{Output layer} & \multicolumn{1}{l}{Batch size} & \multicolumn{1}{l}{Learning rate} & \multicolumn{1}{l}{Epochs} & \multicolumn{1}{l}{$N$} \\ \midrule
1 & t+1 & 1024 & $5\times 10^{-4}$ & 90 & 44562 \\
2 & t+2 & 1024 & $5\times 10^{-4}$ & 90 & 41120 \\
3 & \begin{tabular}[c]{@{}c@{}}t+1, t+2,\\ t+3, t+5\end{tabular} & 248 & $8\times 10^{-4}$ & 90 & 14217 \\
4 & \begin{tabular}[c]{@{}c@{}}t+1, t+2,\\ t+3, SJR\end{tabular} & 248 & $5\times 10^{-4}$ & 90 & 8978 \\
5 & t+1 & 500 & $5\times 10^{-4}$ & 90 & 44562 \\
6 & t+1, t+2, SJR & 500 & $8\times 10^{-4}$ & 90 & 13157 \\
7 & t+1, t+2 & 500 & $8\times 10^{-4}$ & 90 & 41120 \\
8 & t+1, t+2, SJR & 500 & $8\times 10^{-4}$ & 90 & 13157 \\ \bottomrule
\end{tabular}
\caption*{\centering \small \textbf{Table 8. Training steps and settings}. Output layer refers to the outcomes that the neural network was trained to predict (`t+k' refers to citations k-years post-publication, and SJR refers to the publication's SJR value). $N$ refers to the number of observations that we trained on for each step (which is equal to the number of examples for which we had all relevant observations for in our training set).}
\end{table}

\section*{Appendix G: Input data}
\label{sec:AppendixG}

We train our DNN to develop to predict a variety of bibliometric-related and publication-related outcomes based on a large number of author- and publication-level features, which were encoded as a $269 \times 1$ input vector. The input-layer features are described in \hyperref[tab:table9]{Table 9}.

\begin{table}[h!]
\centering
\small
\begin{tabular}{@{}lll@{}}
\toprule
\textbf{Variable}                                               & \textbf{Included for which authors} & \textbf{Lags} \\ \midrule
\textit{h}-index                                          & Up to 20 authors                    & 4 years       \\
Number of publications                                         & Up to 20 authors                    & 4 years       \\
Number of citations received (excluding self-citations)         & Up to 20 authors                    & 4 years       \\
Total grant funding received by authors’ institution            & All                                 & 5 years       \\
Ranking of institution                   & Up to 20 authors                    & —             \\
Number of authors by institution type (academic, govm, company) & All                                 & —             \\
Number of authors on publication                                               & —                                   & —             \\
Publication date                                                & —                                   & —             \\ \bottomrule
\end{tabular}
\label{tab:table9}
\caption*{\centering \small \textbf{Table 9. Features in input layer}. Variables that make up our input vector. All variables are evaluated at the time of publication. In all, each input vector is of the dimension $1 \times 179$.}
\end{table}
In case there are more than 20 authors on a single publication we select the data for the first 15 authors and the last 5 authors, as commonly, authors are ordered in descending order of their share of contributions, with the exception that supervisory or senior authors are listed last. This ensures that we include both the batch of authors that might have made the largest contributions, as well as the authors who had a supervisory role. 

\section*{Appendix H: Using human capital to predict bibliometric- and publication-related outcomes}
\label{sec:AppendixH}
\begin{figure}[h!]
    \centering
    \includegraphics[width=0.8\textwidth]{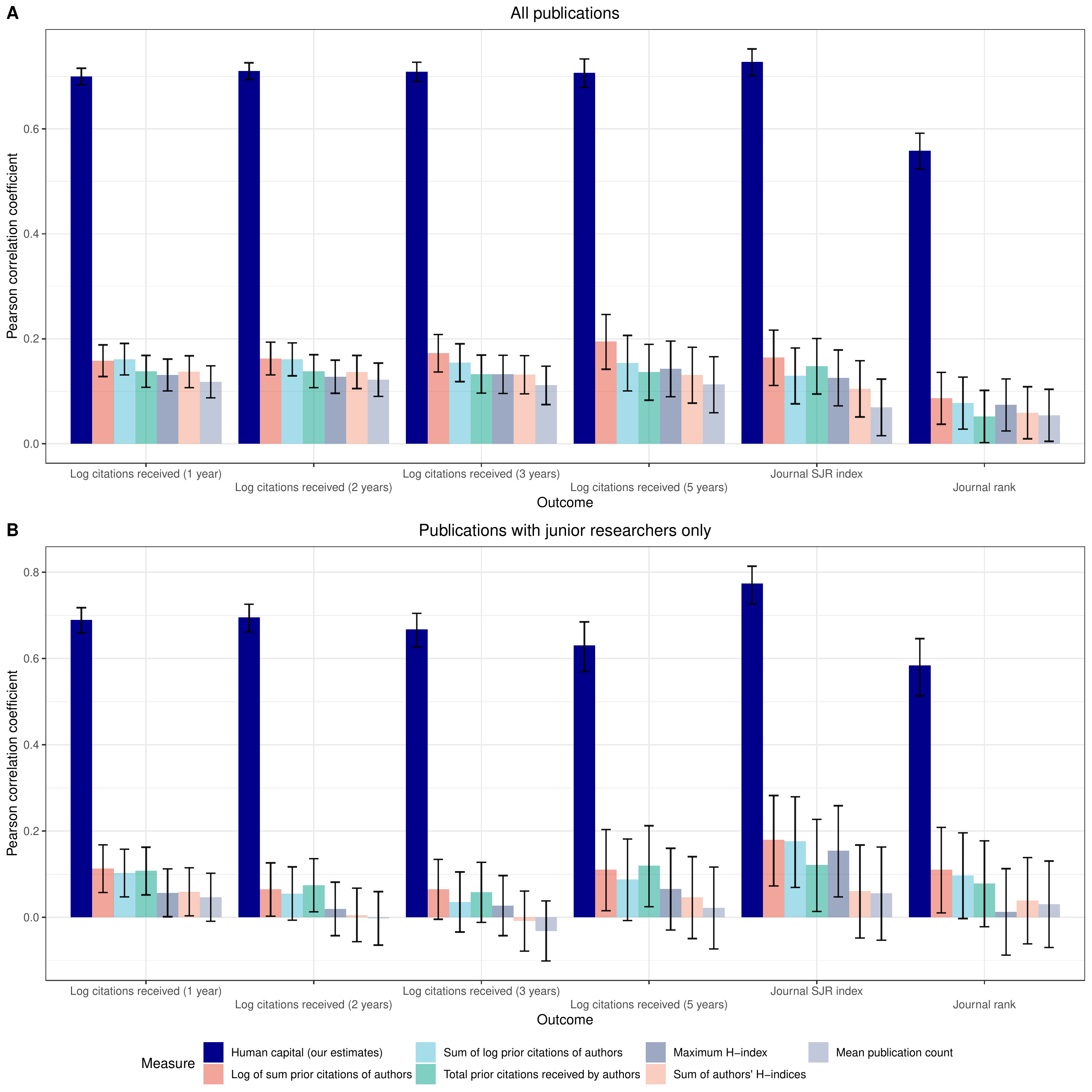}
    \caption*{\centering \small \textbf{Figure 8. Our human capital estimates predict key outcomes on hold-out set} for all publications (top) and publications with junior authors only (bottom). Publications with junior researchers is a subset of the test set consisting of 1,230 publications for which none of the publication authors have more than 2 prior publications.}
    \label{fig:my_label}
\end{figure}

\newpage

\section*{Appendix I: Inputs and performance}
\label{sec:AppendixI}

The computer vision models in our dataset are trained on a wide range of compute budgets. This range spans 4 orders of magnitude for image classification models, and 2 orders of magnitude for object detection models. Training compute is highly variable, yet, with steady growth in the most compute-intensive models (see Figure 3). In particular, the estimated doubling time of computation used in the most compute-intensive models not trained on extra data (for which we have by far the most data) are 8.72 months [95\% CI: 2.22 to 34.33 months] for Image Classification models, and 8.96 months [95\% CI: 2.48 to 32.35 months] for object detection models. This growth rate in compute-intensity is consistent with previous estimates of compute trends in recent years (Sevilla et al., 2022).
\begin{figure}[h!]
\centering
\includegraphics[width=0.9\textwidth]{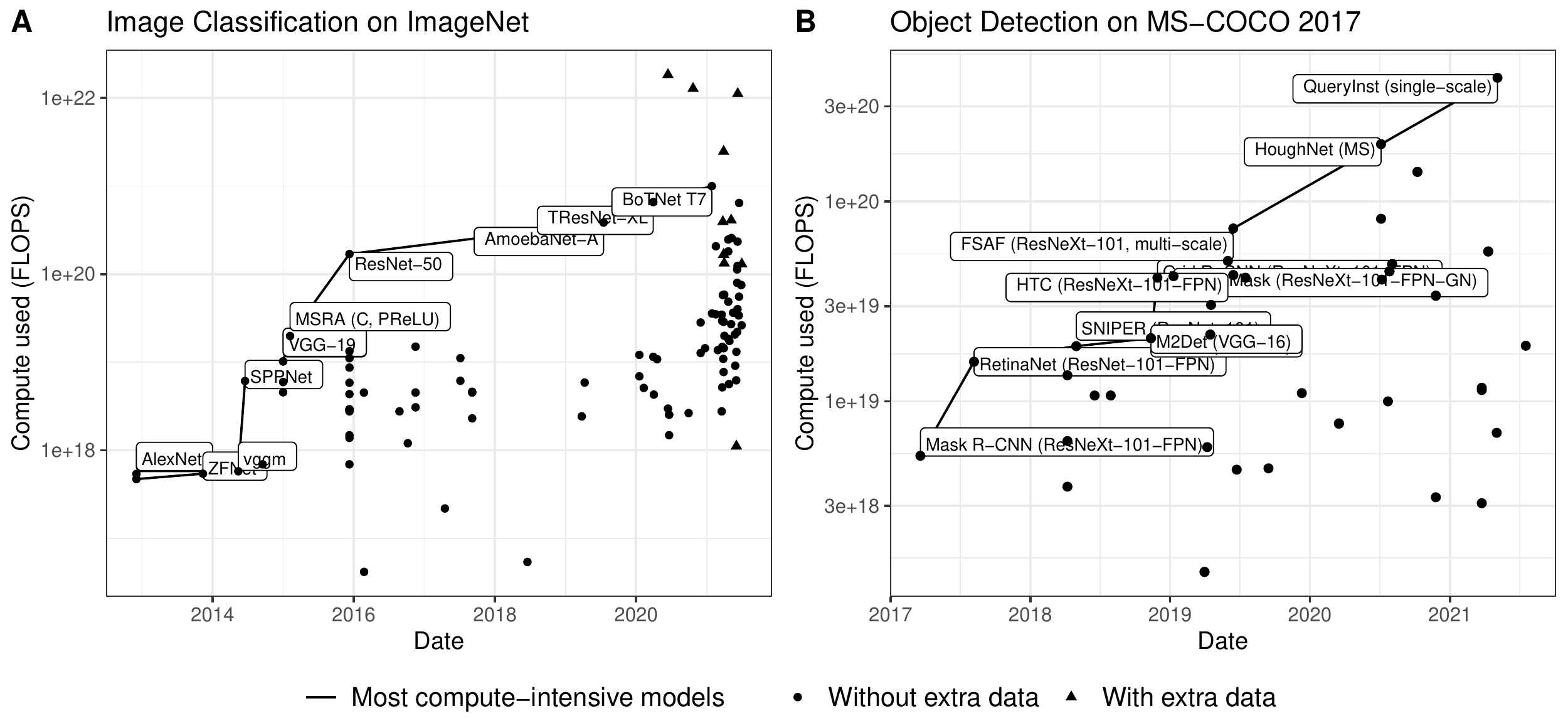}
\caption*{\centering \small \textbf{Figure 9.} Compute intensity of training runs over time for each benchmark. Note: the y-axis is logarithmic ($\log_{10}$).}
\end{figure}

Figure 6 plots model performance against both training compute and human capital. We find that computational capital and human capital are both associated with improved model performance. For the machine learning tasks considered we find roughly a power-law relationship between compute and error rates, in line with the experimental results in the literature (\cite{kaplan2020scaling}; \cite{hestness2017deep}). The relationship is particularly evident for ImageNet. 
\begin{figure}[h!]
\centering
\includegraphics[width=0.9\textwidth]{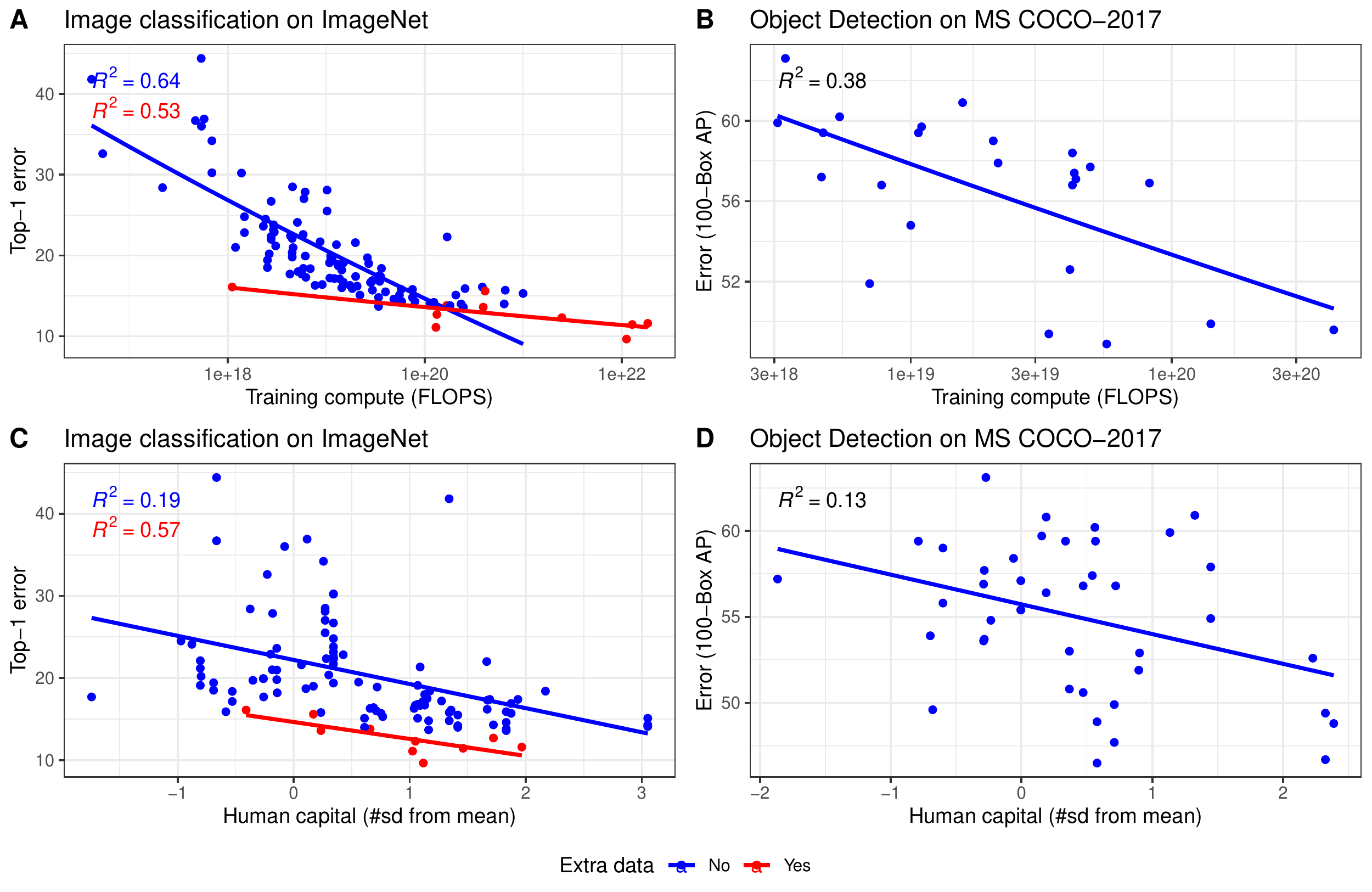}
\caption*{\centering \small \textbf{Figure 10.} Associations between performance and training compute (top panel) and human capital (bottom panel).}
\end{figure}
Additionally, there is some positive association between performance and the estimated amount of human capital, though this association seems generally much weaker than that between computational capital and performance.

\section*{Appendix J: Regression model specifications and estimation procedures}
\label{sec:AppendixJ}

Recall that our empirical estimation is given by:
\begin{equation*}
    \label{eq:14a}
    \log \tilde{g}_{i} = \mathbf{\phi}'\mathbf{1}  + \gamma \log S_{i} + \beta \log C_{i} + \mathbf{\alpha}\mathbf{X} + u_{i}, \hspace{0.15cm} \text{where} \hspace{0.15cm} \mathbf{\phi} \equiv (1-\theta)\begin{bmatrix}
           \log A_1 \\
           \log A_2 \\
           \vdots \\
           \log A_T
         \end{bmatrix}.
\end{equation*}
Table 10 includes each of the full models that we estimate in Section \href{sec:Analysis}{7}.
\newpage

\begin{table}[h!]
\centering
\small
\renewcommand{\arraystretch}{1.6}
\begin{tabular}{@{}lll@{}}
\toprule
Model & Dataset & Specification \\ \midrule
A1 & Image classification & \begin{tabular}[c]{@{}l@{}} $\log \tilde{g}_{i} = \mathbf{\phi}'\mathbf{1}    + \gamma \log S_{i} + \beta \log C_{i} + \alpha_1 \text{extra data}_i + \alpha_2 \text{reimplementation}_i$\\ \end{tabular} \\
A2 & Image classification & \begin{tabular}[c]{@{}l@{}} $\log \tilde{g}_{i} = \mathbf{\phi}'\mathbf{1}    + \gamma \log S_{i} + \beta \log C_{i} + \alpha_1 \text{extra data}_i + \alpha_2 \text{reimplementation}_i$\\$\alpha_3 \text{years from 2012}_i$\end{tabular} \\
B1 & Object detection & \begin{tabular}[c]{@{}l@{}}$\log \tilde{g}_{i} = \mathbf{\phi}'\mathbf{1}    + \gamma \log S_{i} + \beta \log C_{i} + \alpha_1 \text{extra data}_i + \alpha_2 \text{reimplementation}_i$\\ \end{tabular} \\
B2 & Object detection & \begin{tabular}[c]{@{}l@{}}g$\log \tilde{g}_{i} = \mathbf{\phi}'\mathbf{1}    + \gamma \log S_{i} + \beta \log C_{i} + \alpha_1 \text{extra data}_i + \alpha_2 \text{reimplementation}_i$\\ $\alpha_3 \text{years from 2012}_i$\end{tabular} \\
C1 & All computer vision & \begin{tabular}[c]{@{}l@{}}$\log \tilde{g}_{i} = \mathbf{\phi}'\mathbf{1}    + \gamma \log S_{i} + \beta \log C_{i} + \alpha_1 \text{extra data}_i + \alpha_2 \text{reimplementation}_i$\\\end{tabular} \\
C2 & All computer vision & \begin{tabular}[c]{@{}l@{}} $\log \tilde{g}_{i} = \mathbf{\phi}'\mathbf{1}    + \gamma \log S_{i} + \beta \log C_{i} + \alpha_1 \text{extra data}_i + \alpha_2 \text{reimplementation}_i$ \\ $\alpha_3 \text{years from 2012}_i$\end{tabular} \\ \bottomrule
\end{tabular}
\caption*{ \small \centering \textbf{Table 10. Full model specifications}. $\phi'$ denotes the transpose of $\phi$. $\text{reimplementation}_i$ is equal to $1$ if the model $i$ is a re-implementation of prior work, and a $0$ otherwise. $\text{extra data}_i$ is equal to $1$ if model $i$ was trained using extra data (other than the ImageNet training set), and $0$ otherwise (this includes any pre-training of the model, including any of its backbones). The trend coefficients reported in our empirical results correspond to the estimates of the coefficient denoted $\alpha_3$ in this table.}
\end{table}

\begin{table}[h!]
\centering
\small
\renewcommand{\arraystretch}{1.6}
\begin{tabular}{@{}ccccccc@{}}
\toprule
\multicolumn{1}{l}{} & \multicolumn{6}{c}{Model} \\ \midrule
Time-period length & A1 & A2 & B1 & B2 & C1 & C2 \\ \midrule
6 & \begin{tabular}[c]{@{}c@{}}GLS\\ Table 6\end{tabular} & \begin{tabular}[c]{@{}c@{}}GLS\\ Table 12\end{tabular} & \begin{tabular}[c]{@{}c@{}}OLS (HC)\\ Table 6\end{tabular} & \begin{tabular}[c]{@{}c@{}}OLS (HC)\\ Table 12\end{tabular} & \begin{tabular}[c]{@{}c@{}}GLS\\ Table 6\end{tabular} & \begin{tabular}[c]{@{}c@{}}OLS (HC)\\ Table 12\end{tabular} \\
12 & \begin{tabular}[c]{@{}c@{}}GLS\\ Tables 4, 6\end{tabular} & \begin{tabular}[c]{@{}c@{}}OLS (HC)\\ Tables 4, 6\end{tabular} & \begin{tabular}[c]{@{}c@{}}OLS (HC)\\ Tables 4, 6\end{tabular} & \begin{tabular}[c]{@{}c@{}}OLS (HC)\\ Tables 4, 6\end{tabular} & \begin{tabular}[c]{@{}c@{}}OLS (HC)\\ Tables 5, 6\end{tabular} & \begin{tabular}[c]{@{}c@{}}OLS (HC)\\ Tables 5, 6\end{tabular} \\
18 & \begin{tabular}[c]{@{}c@{}}OLS (HC)\\ Table 6\end{tabular} & \begin{tabular}[c]{@{}c@{}}OLS (HC)\\ Table 12\end{tabular} & \begin{tabular}[c]{@{}c@{}}OLS (HC)\\ Table 6\end{tabular} & \begin{tabular}[c]{@{}c@{}}OLS (HC)\\ Table 12\end{tabular} & \begin{tabular}[c]{@{}c@{}}OLS (HC)\\ Table 6\end{tabular} & \begin{tabular}[c]{@{}c@{}}OLS (HC)\\ Table 12\end{tabular} \\ \bottomrule
\end{tabular}
\caption*{\centering \small \textbf{Table 11. Estimation techniques used}. GLS refers to Generalized Least Squares, and OLS (HC) refers to Ordinary Least Squares, with robust covariance matrix estimators with a degrees of freedom correction $(n-1)/(n-k)$ where $n$ is the number of observations and $k$ is the number of explanatory or predictor variables in the model. We use GLS when, after performing a Bruesch-Pagan test for heteroskedastic errors, we reject the null of no heteroskedasticity at a 5\% sign. level.}
\end{table}

\newpage

\section*{Appendix K: Supporting empirical results}
\label{sec:AppendixK}

We re-estimate models A1-C2 with window-lengths 6, 12, and 18 months, and compare our estimates to those obtained in Section \ref{sec:Analysis}. We find that the estimates are mostly similar for all relevant datasets.

\begin{table}[h!]
\centering
\small
\renewcommand{\arraystretch}{1.6}
\begin{tabular}{lccccc}
\toprule
\multicolumn{1}{c}{Data} & \begin{tabular}[c]{@{}c@{}}Time-period\\ length\end{tabular} & Estimates & \multicolumn{2}{c}{Estimates} & \begin{tabular}[c]{@{}c@{}}Log\\ likelihood\end{tabular} \\ \toprule
 &  & \begin{tabular}[c]{@{}c@{}}R\&D elasticity to \\ capital ($\beta$)\end{tabular} & \begin{tabular}[c]{@{}c@{}}R\&D elasticity to \\ human capital ($\gamma$)\end{tabular} & Trend &  \\ \cline{2-6} 
\multirow{3}{*}{\begin{tabular}[c]{@{}l@{}}Image \\classification\end{tabular}} & 6 & \begin{tabular}[c]{@{}c@{}}$\Underset{(0.017)}{0.100}$\signnn\end{tabular} & \begin{tabular}[c]{@{}c@{}}$\Underset{(0.087)}{0.211}$\sign\end{tabular} & \begin{tabular}[c]{@{}c@{}}$\Underset{(0.029)}{0.065}$\end{tabular} & -47.958 \\
 & 12 & \begin{tabular}[c]{@{}c@{}}$\Underset{(0.025)}{0.140}$\signnn\end{tabular} & \begin{tabular}[c]{@{}c@{}}$\Underset{(0.114)}{0.350}$\signn\end{tabular} & \begin{tabular}[c]{@{}c@{}}$\Underset{(0.014)}{0.052}$\signnn\end{tabular} & -39.786 \\
 & 18 & \begin{tabular}[c]{@{}c@{}}$\Underset{(0.021)}{0.154}$\signnn \end{tabular} & \begin{tabular}[c]{@{}c@{}}$\Underset{(0.121)}{0.184}$\nosign \end{tabular} & \begin{tabular}[c]{@{}c@{}}$\Underset{(0.015)}{0.042}$\signn \end{tabular} & -68.709 \\
\multirow{3}{*}{\begin{tabular}[c]{@{}l@{}}Object \\detection\end{tabular}} & 6 & \begin{tabular}[c]{@{}c@{}}$\Underset{(0.089)}{0.210}$\sign \end{tabular} & \begin{tabular}[c]{@{}c@{}}$\Underset{(0.159)}{0.424}$\sign \end{tabular} & \begin{tabular}[c]{@{}c@{}}$\Underset{(0.048)}{-0.014}$ \end{tabular} & -27.169 \\
 & 12 & \begin{tabular}[c]{@{}c@{}}$\Underset{(0.090)}{0.253}$\signn \end{tabular} & \begin{tabular}[c]{@{}c@{}}$\Underset{(0.158)}{0.319}$\nosign\end{tabular} & \begin{tabular}[c]{@{}c@{}}$\Underset{(0.030)}{0.013}$\nosign \end{tabular} & -43.187 \\
 & 18 & \begin{tabular}[c]{@{}c@{}}$\Underset{(0.109)}{0.236}$\sign \end{tabular} & \begin{tabular}[c]{@{}c@{}}$\Underset{(0.197)}{0.376}$\sign \end{tabular} & \begin{tabular}[c]{@{}c@{}}$\Underset{(0.022)}{0.054}$\sign \end{tabular} & -41.862 \\
\multirow{3}{*}{\begin{tabular}[c]{@{}l@{}}Computer \\vision (pooled)\end{tabular}} & 6 & \begin{tabular}[c]{@{}c@{}}$\Underset{(0.015)}{0.130}$\signnn \end{tabular} & \begin{tabular}[c]{@{}c@{}}$\Underset{(0.072)}{0.274}$\signnn  \end{tabular} & \begin{tabular}[c]{@{}c@{}}$\Underset{(0.020)}{0.050}$\sign \end{tabular} & -76.372 \\
 & 12 & \begin{tabular}[c]{@{}c@{}}$\Underset{(0.016)}{0.142}$\signnn \end{tabular} & \begin{tabular}[c]{@{}c@{}}$\Underset{(0.068)}{0.263}$\signnn \end{tabular} & \begin{tabular}[c]{@{}c@{}}$\Underset{(0.005)}{0.030}$\signnn \end{tabular} & -80.164 \\
 & 18 & \begin{tabular}[c]{@{}c@{}}$\Underset{(0.018)}{0.144}$\signnn \end{tabular} & \begin{tabular}[c]{@{}c@{}}$\Underset{(0.079)}{0.207}$\signn \end{tabular} & \begin{tabular}[c]{@{}c@{}}$\Underset{(0.004)}{0.020}$\signnn\end{tabular} & -98.355 \\ \bottomrule
\end{tabular}
\caption*{\centering \small \textbf{Table 12. Estimation results for separate models with alternative window lengths}. Estimates of models A2, B2, and C2 with different window-lengths. Specifications are the same as in the main analysis in Section} \ref{sec:Analysis}.
\label{tab:tab11}
\end{table}

\end{document}


\section{Appendix A: Theory results}

 Using (1-3), we can derive the rates at which capital and ideas grow:
\begin{equation}
    g_k(t) \equiv \frac{\dot{K}(t)}{K(t)} = c_k \bigg[\frac{A(t) L(t)}{K(t)}\bigg]^{1-\alpha} - \delta, \hspace{0.15cm} \text{where} \hspace{0.15cm} c_k \equiv s(1-\alpha_k)^\alpha (1-\alpha_l)^{1-\alpha}, 
\end{equation}
\begin{equation}
    g_a(t) \equiv \frac{\dot{A}(t)}{A(t)} = c_a K(t)^\beta L(t)^\gamma A(t)^{\theta-1}, \hspace{0.15cm} \text{where} \hspace{0.15cm}  c_a \equiv B \alpha_k^\beta \alpha_l^\gamma.
\end{equation}
Along the balanced growth path (defined as an equilibrium path where $Y(t)$, $K(t)$, $A(t)$ and $L(t)$ grow at a constant rate), it can be shown that:
\begin{equation}
   \tilde{g}_k(t) \equiv \frac{\dot{g}_k(t)}{g_k(t)} = (1-\alpha)(g_a + n - g_k),  \hspace{0.15cm} \text{and} \hspace{0.15cm} \tilde{g}_a(t) \equiv \frac{\dot{g}_a(t)}{g_a(t)} = \beta g_k + \gamma n - (1-\theta) g_a.
\end{equation}
The steady-state rates of growth in ideas and capital can then simply be found by solving for $g_k(t)$ and $g_a(t)$ that solves $\tilde{g}_k(t) = \tilde{g}_a(t) = 0$. Solving this system yields the following equilibrium growth rates (equilibrium growth rates are marked with the ${}^*$ superscript):
\begin{equation}
    g^{*}_a =\frac{\beta + \gamma}{1-\beta -\theta}n, \hspace{0.15cm} \text{and} \hspace{0.15cm}  g^{*}_k = \frac{1-\theta + \gamma}{1-\beta -\theta}n.
\end{equation}
which can be shown to be unique and stable.

\subsubsection{Proof of proposition 1.}

Let $\Delta_X$ denote $X'-X$. By assumption, $\Delta_\beta \geq - \Delta_\gamma \geq 0$. The steady-state rate of growth in ideas is increased iff
\begin{equation}
    \frac{\beta' + \gamma'}{1-\beta' -\theta}  \geq \frac{\beta + \gamma}{1-\beta-\theta}.
\end{equation}
which follows from the fact that $\Delta_\beta \geq - \Delta_\gamma \geq 0$. To prove that the steady-state rate of economic growth is also increased, it suffices to show that the steady-state rate of capital accumulation is not decreased. To do so, note that this is true if,
\begin{equation}
\frac{1-\theta+\gamma'}{1-\beta'-\theta} \geq  \frac{1-\theta+\gamma}{1-\beta-\theta}
\end{equation}
Rearranging, this yields the condition:
\begin{equation}
    -\Delta_\gamma \leq A\Delta_\beta, \hspace{0.15cm} \text{where} \hspace{0.15cm} A \equiv \frac{1-\theta + \gamma}{1-\beta -\theta} \geq 1,
\end{equation}
which is true on the assumption $-\Delta_\gamma \leq \Delta_\beta$. \qedsymbol{}

\section{Consistency of Feasible Within-Time Estimator}
The following is a rough argument that our Feasible Within-Time Estimator is consistent. Suppose the production function is described as follows:
\begin{equation}
\dot{a}_i = \gamma s_i + \beta c_i + \theta A_{t} + u_{i,t},
\end{equation}
and the following assumptions hold:
\begin{enumerate}
    \item $A_t$ has additively separable variance-components: $A_t=A_t^* + \delta_t$, where $A_t^*$ and $\delta_t$ are independent shocks are realised in time $t$,
    \item The idiosyncratic shock $A_t^*$ becomes known only after the research team decides the amount of inputs to employ at period $t$ and is not predictable ex-ante.
\end{enumerate}
We can then show that the feasible Within Time Estimator described in Section 3.2. is consistent for $\gamma$ and $\beta$ whenever the $\delta_t$ is a linear trend, i.e. $\delta_t = \delta t$, and observations are uniformly distributed over time. To see this, notice that the Within-Time transformed equation at time $T$ is
\begin{equation}
       \big(\dot{a}_i-\overline{\dot{a}_i}\big)= \gamma  \big(s_{i}  -\overline{s}_{T} \big) + \beta  \big(c_{i} -\overline{c}_{T} \big) + \theta\big(A_T^* + \delta T - \frac{1}{n}\sum_{t = T-l }^{t= T+l}(A_t^* + \delta t)\big) + \big(u_{i,T} - \overline{u}_{i,T}\big)
\end{equation}
Observations being uniformly distributed over time implies that $\frac{1}{n}\sum_{t = T-l }^{t= T+l}\delta t \simeq \delta T$, hence, on this assumption, the Within-Time transformed equation at time $T$ becomes:
\begin{equation}
       \big(\dot{a}_i-\overline{\dot{a}_i}\big)= \gamma  \big(s_{i}  -\overline{s}_{T} \big) + \beta  \big(c_{i} -\overline{c}_{T} \big) + \theta \big(A_T^* -\overline{A_t^*}\big) + \big(u_{i,T} - \overline{u}_{i,T}\big),
\end{equation}
which successfully eliminates the unobserved heterogeneous component $\delta_t$.

\section{Appendix B: Data-set construction}

In this section, we describe how each of our data-sets were constructed. We provide a high-level overview of our methodology. For full details, see the \underline{\href{https://docs.google.com/document/d/1xZWALkldodSipA_qA1lsSaANrEo083QtD3nW4Z9BfqM/edit?usp=sharing}{online appendix}}.

\subsection{Data on Deep Learning models}

[To do]

\subsection{Data on researcher human capital}

Our model for human capital draws on data from a number of sources, which are described below.

\subsubsection{Bibliometric panel dataset}

[To do: insert definitions for DL, L distances]
[To do: insert all citations]
[To do: insert Table for key words used in matching]
[To do: consider a table or graph summarizing this section? Maybe: step $i$, number of authors found, error rate?]

We generate a yearly panel data-set of feature for 223,703 authors on 115,235 Machine Learning papers from 1993 to 2019 using data from arXiv, a popular open-access repository for Machine Learning papers; Microsoft Project Academic Knowledge; and Scopus (TODO: cite all). Our panel dataset contains yearly series on all of the following author-specific features: number of publications, number of total citations, and authors’ h-index (a measure of researcher productivity and impact, see Hirsh, 2005 (TODO cite)). We first describe our matching procedure and then the generation of our panel dataset.

To build the data-set, we query arXiv for all papers under the following categories: stat.ML, cs.AI, cs.CL, cs.CV, and cs.LG to obtain the entire universe of Machine Learning papers from 1993 to 2019 [TODO: explain why we start with arXiv, popularity in ML, etc.]. Using the procedure described in \cite{muller_is_2020}, we match these papers to their corresponding entries in the Microsoft Project Academic Knowledge database to obtain author names and affiliations. We then begin iterations of computationally more intensive matching procedures to match the papers to Scopus, performing sampling checks along the way that our matches are accurate. First, we match 30,463 papers using DOI data in the Microsoft dataset. Second, 28,728 additional papers are matched by title using the Damerau-Levenshtein distance, a string-distance measure. A random sample of 50 papers matches 96\%, with a false negative rate of 4\%. Third, we search for authors in Scopus using exact string matches. Due to the commonality of some author names, we download the first 26 papers for each result in Scopus. We check if the authors are Machine Learning researchers by querying papers' abstracts for common words in Table [TODO: insert]. A random sample of 50 matches has a 40\% false negative rate and 4\% false positive rate. This produces an additional 24,954 authors. Fourth, for each remaining author name in the Microsoft dataset, we download a sample of 100 papers from Microsoft and from Scopus for each author in our dataset. We then match paper titles in these samples to identify common authors. To avoid selecting coauthors, we match last names either exactly (for names fewer than 3 characters) or if the Damerau-Levenshtein distance is less than 0.5 (for names longer than 3 characters). On a random sample of 100 matches, we have a 100\% success rate. Along with matching authors for the DOI-matched papers, this nets an additional 47,308 authors. Fifth, we repeat the same procedure but instead of querying Scopus by author name, we query Scopus for each author's entire set of \textit{publications}. We repeat the same procedure as in step 4 above to match authors, matching an additional 32,913 authors. On a random sample of 50 matches, we obtain a false positive rate of 4\%.

Now we describe our procedure for generating panel data. To generate each series, we used bibliometrix, an R tool (Aria and Corrado, 2017 cite), to query Scopus’ API and retrieve all the publications of all author dataset. For each publication, we then used pybliometrics, a Python interface to Scopus (Rose and Kitchin, 2019 TODO cite), to retrieve the citation trajectories—the number of incoming citations—for each publication per year, excluding self-citations. Author’s research fields were identified by retrieving Scopus’ subject area data on each of their publications. For each author, we further gathered data on fractional credit scores for citations and publications, which were generated by dividing the number of citations and publications by the number of authors on each publication.

\subsubsection{Grants dataset}

Data on grants received by institutions comes from the Dimensions database of academic publications (TODO: cite). We query the API for each institution in the Scopus panel dataset on author affiliations. This returns nominal data in USD on all grants per project-year and per institution-year. In order to make intertemporal comparisons, we deflate the institution-level data using the GDP implicit price deflator (TODO: FRED, 2021). We use the Damerau-Levenshtein measure of string distance to match the institution-year data to the matched Scopus-Microsoft cross-sectional paper titles dataset. For the top 1000 papers, we find 731 matches, of which 1.2\% are false positives. Overall, we match 50.8\% of institutions in our titles dataset. On a random sample of 100 matches, 5\% were found to be false positives. We report total grants received for the past 5 years by all authors’ departments or employers in real 2015 USD.

\subsubsection{Institutional rankings data}

Institutional rankings for Computer Science was generated using Computer Science publication data by csmetrics.org (TODO cite), an online dataset of institutional publication and citation metrics for computer science. This data is based on measured (retrospective) and predictive (prospective) metrics to compute a measure of publication impact of computer science publications by institution.  This institution in our dataset was matched to the entities in the csmetrics.org database by using Levenshtein distance, a measure of string-distance. Since not all institutions in our dataset were also present in csmetrics.org’s database, we were able to assign ranks to only 89.6\% of unique institutions in our dataset. The institutions we were unable to retrieve rankings for were smaller institutions that did not appear often in our data. In particular, institutional rankings were assigned in 99.4\% of authors for whom institutional data was available. Moreover, the matching did not produce false-matches: of a random sample of 120 matches of institutions, only 0.9\% of matches were found to be incorrect.

\subsubsection{Computer science publication venue rankings data}

Data on yearly rankings of Computer Science publication venues was retrieved from SCImago Journal \& Country Rank (TODO: cite), a portal for scientific indicators developed from the information contained in the Scopus database. In particular, we use their SJR indicator, which ranks scholarly journals, conferences and proceedings based on citation weighting schemes and eigenvector centrality to quantify the impact and prestige of publication venues (TODO: González-Pereira et al., 2010). Their database contains 5147 unique Computer Science venues spanning the 1999-2020 period.  We used the Levenshtein distance, a measure of string distance, to match the journal in SCImago’s database to the journals in which the articles dataset were published in. Overall, we were able to assign 15,000 of publications with journal rankings and SJR indicators. In case rankings were missing for the exact year of publication, we assigned each journal the rank of either its rank in the year prior to publication, or its rank in the year post-publication. If asynchronous rankings one-year apart were not available, we considered two-years prior and post-publication. Only 0.6\% of ranking assignments were made with asynchronous rankings of the kind described. Of a random sample of 100 publications, we found no incorrect matches between the Scopus-reported publication venue and the entity in SCImago’s database to which it was matched, suggesting our matching was performed highly accurately.

\section{Appendix C: Model training}

\subsection{Training Procedure and hyper-parameter settings}

[To do]

\begin{itemize}
    \item Gelu Activation function
    \item Learning rates
    \item Batch-sizes
    \item Loss-weights
\end{itemize}

\section{Appendix D: Human Capital Estimates}
By computing the activation of the human-capital unit in our model, we are able to infer the total quality-adjusted research input for any given publication. We can further examine how this `estimate' relates to various publication-related outcomes.
\begin{figure}[h!]
    \centering
    \textbf{Table 2: Examples of publications and the associated human capital unit activations}\par\medskip
    \footnotesize
\begin{tabular}{lllllll}
\hline
Publication title & Authors & Date published & \begin{tabular}[c]{@{}l@{}}Human capital \\ activation\\ (standard dev.\\ from mean)\end{tabular} & \begin{tabular}[c]{@{}l@{}}Average\\ H-index \\ author\end{tabular} & \begin{tabular}[c]{@{}l@{}}Citations\\ after one \\ year\end{tabular} & \begin{tabular}[c]{@{}l@{}}Journal/\\ conference\end{tabular} \\ \hline
\begin{tabular}[c]{@{}l@{}}FBNet: Hardware-Aware \\ Efficient ConvNet Design\\ via Differentiable Neural \\ Architecture Search\end{tabular} & Wu et al. & 9 Dec 2018 & 3.83 &  &  &  \\
\begin{tabular}[c]{@{}l@{}}Rethinking the Inception \\ Architecture for Computer \\ Vision\end{tabular} & Szegedy et al. & 2 Dec 2015 & 3.82 &  &  &  \\
\begin{tabular}[c]{@{}l@{}}Feature Denoising for\\  Improving Adversarial \\ Robustness\end{tabular} & Xie et al. & 9 Dec 2018 & 3.78 &  &  &  \\ \hline
\end{tabular}
    \caption*{}
    \label{fig:my_label}
\end{figure}

We find that the activation of the `human-capital unit' is strongly positively correlated with bibliometric and publication-related outcomes. 

\section{Models estimated}

\section{Appendix E: Additional figures}

[To do]

\begin{figure}[h!] 
    \centering
    \textbf{Figure 2: Predicted and ground-truth values on test-set}\par\medskip
   \includegraphics{dnn_performance.pdf}
    \caption*{}
    \label{fig:my_label}
\end{figure}

\begin{figure}[h!]
    \centering
    \textbf{Figure: Human capital activations and publication outcomes}\par\medskip
    \includegraphics[scale=0.5]{human_capital_stdv.pdf}
    \caption*{\small Figure. Human capital unit activations (in number of standard deviations from mean activation) plotted against bibliometric and publication-related outcomes.}
    \label{fig:my_label}
\end{figure}

\section{Appendix F: Machine Learning models used}

[To do]


\section{Appendix}
\subsection{Theory—derivations}

Using (1-3), we can derive the rates at which capital and ideas grow:
\begin{equation}
    g_k(t) \equiv \frac{\dot{K}(t)}{K(t)} = c_k \bigg[\frac{A(t) L(t)}{K(t)}\bigg]^{1-\alpha} - \delta, \hspace{0.15cm} \text{where} \hspace{0.15cm} c_k \equiv s(1-\alpha_k)^\alpha (1-\alpha_l)^{1-\alpha}, 
\end{equation}
\begin{equation}
    g_a(t) \equiv \frac{\dot{A}(t)}{A(t)} = c_a K(t)^\beta L(t)^\gamma A(t)^{\theta-1}, \hspace{0.15cm} \text{where} \hspace{0.15cm}  c_a \equiv B \alpha_k^\beta \alpha_l^\gamma.
\end{equation}
Along the balanced growth path (defined as an equilibrium path where $Y(t)$, $K(t)$, $A(t)$ and $L(t)$ grow at a constant rate), it can be shown that:
\begin{equation}
   \tilde{g}_k(t) \equiv \frac{\dot{g}_k(t)}{g_k(t)} = (1-\alpha)(g_a + n - g_k),  \hspace{0.15cm} \text{and} \hspace{0.15cm} \tilde{g}_a(t) \equiv \frac{\dot{g}_a(t)}{g_a(t)} = \beta g_k + \gamma n - (1-\theta) g_a.
\end{equation}
The steady-state rates of growth in ideas and capital can then simply be found by solving for $g_k(t)$ and $g_a(t)$ that solves $\tilde{g}_k(t) = \tilde{g}_a(t) = 0$. Solving this system yields the following equilibrium growth rates (equilibrium growth rates are marked with the ${}^*$ superscript):
\begin{equation}
    g^{*}_a =\frac{\beta + \gamma}{1-\beta -\theta}n, \hspace{0.15cm} \text{and} \hspace{0.15cm}  g^{*}_k = \frac{1-\theta + \gamma}{1-\beta -\theta}n.
\end{equation}
which can be shown to be unique and stable. [Add transition diagram that shows effect of $\beta$-shock diagrammatically.]

\subsubsection{Proof of proposition 1.}

Let $\Delta_X$ denote $X'-X$. By assumption, $\Delta_\beta \geq - \Delta_\gamma \geq 0$. The steady-state rate of growth in ideas is increased iff
\begin{equation}
    \frac{\beta' + \gamma'}{1-\beta' -\theta}  \geq \frac{\beta + \gamma}{1-\beta-\theta}.
\end{equation}
which follows from the fact that $\Delta_\beta \geq - \Delta_\gamma \geq 0$. To prove that the steady-state rate of economic growth is also increased, it suffices to show that the steady-state rate of capital accumulation is not decreased. To do so, note that this is true if,
\begin{equation}
\frac{1-\theta+\gamma'}{1-\beta'-\theta} \geq  \frac{1-\theta+\gamma}{1-\beta-\theta}
\end{equation}
Rearranging, this yields the condition:
\begin{equation}
    -\Delta_\gamma \leq A\Delta_\beta, \hspace{0.15cm} \text{where} \hspace{0.15cm} A \equiv \frac{1-\theta + \gamma}{1-\beta -\theta} \geq 1,
\end{equation}
which is true on the assumption $-\Delta_\gamma \leq \Delta_\beta$. \qedsymbol{}

\subsection{Appendix: Further details on data collection}

\subsubsection{Data on Deep Learning models}

\subsubsection{Data on researcher human capital}

\subsubsection{Bibliometric panel dataset}

We generated a yearly panel dataset of features for each of the 223,703 authors in our Author dataset using data from Scopus, Elsevier’s abstract, indexing, and citation database. Our dataset covers all the authors we were able to identify in the dataset from McIntyre et al. (cite), which includes 115,235 paper entries from Microsoft Project Academic Knowledge and arXiv from 1993 to 2019. Our panel dataset contains yearly series on all of the following author-specific features: number of publications, number of total citations, and authors’ h-index (a measure of researcher productivity and impact, see Hirsh, 2005 (cite)). We first describe our matching procedure and then the generation of our panel dataset.

[To do: This is a bit messy, figure out how to describe better!]

We begin with matching all papers in the dataset from McIntyre et al. (cite) (hereafter MMA dataset) to their entries in the Scopus dataset, to enable us to subsequently identify all its authors. First, we identify all the papers from this dataset for which the MMA dataset has DOI data. In total, 38,394 papers were identified, of which 30,463 are found in Scopus. For the papers without DOIs, we attempt matching these to those in the Scopus database using inexact title matching. We then perform an inner join with the MS titles using the Damerau-Levenshtein distance. See Table 1 for summary statistics. We test the algorithm by manually coding a random sample of 50 papers from the MS dataset. 25 papers were found in Scopus, of which the inner join matched 23, for a false negative rate of 8\%. A second random sample was performed on 50 papers, finding 25/25 papers. This inexact matching approach enabled us to identify a further 28,728 papers, for a total of 51.4\% of the MS titles.

The MMA paper dataset contains 161,644 unique author IDs. Given the above paper data, we set aside Scopus author entries for which we already have paper data. This is 93,433 authors in MS and 149,568 in Scopus. Note that 3.54\% of papers in MS have more authors than in Scopus and 11.61\% have more authors in Scopus. I include these 94,248 additional authors from Scopus in the dataset as well. This discrepancy may be due partly to different drafts of papers in the two datasets or to erroneous paper matching between the two datasets.

Next, we match the authors in our dataset to the corresponding entries in the Scopus dataset  using their full names, as given in the MMA dataset. Since multiple authors will have the same names, we filter the results to ensure that there is a match between authors in the MMA dataset and Scopus query results. To do this, we first query Scopus using the first 26 papers for the first Scopus entry for each query. We opt for the first author to reduce time querying the API. Then, we filter the entries by exact matching the strings in Table X on the papers’ abstracts. Finally, because some author names in the MMA dataset return multiple profiles in Scopus and because some authors in the former dataset are not unique, we only consider injective matches between the datasets. In a random sample of 50 authors, 28 authors were found, 2 of which were false positives, for an error rate of 4\%. This process found 36,719 authors. Filtering for injective matches produces 27,337 authors. We then repeat the above, querying Scopus for the remaining of authors. This produces an additional 24,954 entries, again filtering the entries by exact matching the strings in Table X on the papers’ abstracts and considering only injective matches.

Next, we attempt matching authors from the McIntyre et al. dataset to authors in Scopus dataset using paper data. For each of the remaining authors, we download 100 papers from MS and from Scopus. For each MS author, we iterate through the entire set of Scopus authors and count the number of paper titles that are an exact match between authors. Denote this set as $S$. For $s\in\{\text{Scopus authors}\}$, if $s=\text{argmax}(S)$ and $|s| = 1$, we consider $s$ a potential match. Then, to ensure that the algorithm is not selecting a coauthor, we ensure that either the last names of the authors have 3 or fewer letters and are exact matches or have greater than 3 letters and the string distance between the last names is less than or equal to $0.5*\floor(\text{“MMA last name”})$. If either of these conditions are satisfied, we consider s to be a match. Sampling and manually verifying 100 random authors gives a 0\% error rate. I match an additional 16,702 authors. As some authors are now removed, we iterate again through the above procedure, matching an additional 30,606 authors. Altogether, we are now at 108,891 authors.

Finally, we consider a paper-first approach. We download the entire set of papers (ML and non-ML) for each remaining author in MS. To cut down on computation time, we first sample a random set of 5 papers from each author. For each paper, we search the exact title match in Scopus. If there is a match, we first calculate whether the string distance between the titles is $≤ 5$. If not, it does not match the titles. If the titles are a close match, we then choose the author in Scopus that has the minimum DL distance to the author in MS. Finally, as a sanity check, we ensure that the string distance between the matched names is less than or equal to $0.5*\floor(\text{“MMA name”})$. We match 30,653 additional authors. We repeat the procedure but with the entire set of papers for each author in MS, rather than a random sample of 5. In addition, a few authors had hyphenated last names, which were occasionally coded as middle names and other times as last names. So we concatenate middle and last names before performing matching based on string distance. We find an additional 2,260 authors. To determine the error rate, we randomly sample 50 authors and manually check the algorithm. It returns 1 false positive, for an error rate of 4\%. Note that in the random sample of 50, the sets of papers for 9 MS authors mapped to more than one author in Scopus. We are not sure which dataset is at fault. However, we can say that we are at the level of fidelity of the original data. We were unable to match 22,056 authors from arXiv. Figure X displays the steps of the algorithm along with authors matched at each step and error rates, and Figure X displays the intersection of the MS and Scopus author datasets.

Now we describe our procedure for generating panel data. To generate each series, we used bibliometrix, an R tool (Aria and Corrado, 2017 cite), to query Scopus’ API and retrieve all the publications of all author dataset. For each publication, we then used pybliometrics, a Python interface to Scopus (Rose and Kitchin, 2019 cite), to retrieve the citation trajectories—the number of incoming citations—for each publication per year, excluding self-citations. Author’s research fields were identified by retrieving Scopus’ subject area data on each of their publications. For each author, we further gathered data on fractional credit scores for citations and publications, which were generated by dividing the number of citations and publications by the number of authors on each publication.

\subsubsection{Grants dataset}

Data on grants received by institutions comes from Dimensions’ API (cite), a database of academic publications. We query the API for each institution in the Scopus panel dataset on author affiliations. This returns nominal data in USD on all grants per project-year and per institution-year. In order to make intertemporal comparisons, we deflate the institution-level data using the GDP implicit price deflator (FRED, 2021 cite). We use the Damerau-Levenshtein distance measure of string distance to match the institution-year data to the matched Scopus-Microsoft cross-sectional paper titles dataset. For the top 1000 papers, we find 731 matches, of which 1.2\% are false positives. Overall, we match 50.8\% of institutions in our titles dataset. On a random sample of 100 matches, 5\% were found to be false positives. We report total grants received for the past 5 years by all authors’ departments or employers in real 2015 USD.

\subsubsection{Institutional rankings data}

Institutional rankings for Computer Science was generated using Computer Science publication data by csmetrics.org (cite), an online institutional publication and citation metrics for computer science. This data is based on measured (retrospective) and predictive (prospective) metrics to compute a measure of publication impact of computer science publications by institution. 

This institution in our dataset was matched to the entities in the csmetrics.org database by using Levenshtein distance, a measure of string-distance. Since not all institutions in our dataset were also present in csmetrics.org’s database, we were able to assign ranks to only 89.6\% of unique institutions in our dataset. The institutions we were unable to retrieve rankings for were smaller institutions that did not appear often in our data. In particular, institutional rankings were assigned in 99.4\% of authors for whom institutional data was available. Moreover, the matching did not produce false-matches: of a random sample of 120 matches of institutions, only 0.9\% of matches were found to be incorrect.

\subsubsection{Computer science publication venue rankings data}

Data on yearly rankings of Computer Science publication venues was retrieved from SCImago Journal & Country Rank (cite), a portal for scientific indicators developed from the information contained in the Scopus database. In particular, we use their SJR indicator, which ranks scholarly journals, conferences and proceedings based on citation weighting schemes and eigenvector centrality to quantify the impact and prestige of publication venues (González-Pereira et al., 2010 cite). Their database contains 5147 unique Computer Science venues spanning the 1999-2020 period. 

We used the Levenshtein distance, a measure of string distance, to match the journal in SCImago’s database to the journals in which the articles dataset were published in. Overall, we were able to assign x\% of publications with journal rankings and SJR indicators. In case rankings were missing for the exact year of publication, we assigned each journal the rank of either its rank in the year prior to publication, or its rank in the year post-publication. If asynchronous rankings one-year apart were not available, we considered two-years prior and post-publication. Only y\% of ranking assignments were made with asynchronous rankings of the kind described. Of a random sample of 100 publications, we found no incorrect matches between the Scopus-reported publication venue and the entity in SCImago’s database to which it was matched, suggesting our matching was performed highly accurately.

\newpage
\section{Appendix: Further details on human capital model training procedure}

\subsection{Training Procedure and hyper-parameter settings}

\begin{itemize}
    \item Gelu Activation function
    \item Learning rates
    \item Batch-sizes
    \item Loss-weights
\end{itemize}
\subsubsection{Human Capital Estimates}
By computing the activation of the human-capital unit in our model, we are able to infer the total quality-adjusted research input for any given publication. We can further examine how this `estimate' relates to various publication-related outcomes.
\begin{figure}[h!]
    \centering
    \textbf{Table 2: Examples of publications and the associated human capital unit activations}\par\medskip
    \footnotesize
\begin{tabular}{lllllll}
\hline
Publication title & Authors & Date published & \begin{tabular}[c]{@{}l@{}}Human capital \\ activation\\ (standard dev.\\ from mean)\end{tabular} & \begin{tabular}[c]{@{}l@{}}Average\\ H-index \\ author\end{tabular} & \begin{tabular}[c]{@{}l@{}}Citations\\ after one \\ year\end{tabular} & \begin{tabular}[c]{@{}l@{}}Journal/\\ conference\end{tabular} \\ \hline
\begin{tabular}[c]{@{}l@{}}FBNet: Hardware-Aware \\ Efficient ConvNet Design\\ via Differentiable Neural \\ Architecture Search\end{tabular} & Wu et al. & 9 Dec 2018 & 3.83 &  &  &  \\
\begin{tabular}[c]{@{}l@{}}Rethinking the Inception \\ Architecture for Computer \\ Vision\end{tabular} & Szegedy et al. & 2 Dec 2015 & 3.82 &  &  &  \\
\begin{tabular}[c]{@{}l@{}}Feature Denoising for\\  Improving Adversarial \\ Robustness\end{tabular} & Xie et al. & 9 Dec 2018 & 3.78 &  &  &  \\ \hline
\end{tabular}
    \caption*{}
    \label{fig:my_label}
\end{figure}

We find that the activation of the `human-capital unit' is strongly positively correlated with bibliometric and publication-related outcomes. In particular, we find that one standard deviation increase in human capital is associated with an X\% increase in the number of citations received.

\section{Additional figures}

\begin{figure}[h!]
    \centering
    \textbf{Figure 2: Predicted and ground-truth values on test-set}\par\medskip
    \includegraphics{dnn_performance.pdf}
    \caption*{}
    \label{fig:my_label}
\end{figure}

\begin{figure}[h!]
    \centering
    \textbf{Figure: Correlations for Machine Learning Models}\par\medskip
    \includegraphics[scale=0.5]{correlations.pdf}
    \caption*{\small }
    \label{fig:my_label}
\end{figure}

\begin{figure}[h!]
    \centering
    \textbf{Figure: Human capital activations and publication outcomes}\par\medskip
    \includegraphics[scale=0.5]{human_capital_stdv.pdf}
    \caption*{\small Figure. Human capital unit activations (in number of standard deviations from mean activation) plotted against bibliometric and publication-related outcomes.}
    \label{fig:my_label}
\end{figure}

\subsection{The relationship between $\log(S)$ and $Z(S)$}

We assume that $S$ can be described by a log-normal distribution. [Provide reasons]. Given this assumption, the z-score of $S$, $Z(S)$, will be equivalent to to $\log(S)$ up to an affine transformation. Obviously, by definition:
\begin{equation}
    S \sim \text{Lognormal}(\mu_1, \sigma_1) \implies  \log(S) \sim \text{N}(\mu_2, \sigma_2),
\end{equation}
and therefore:
\begin{equation}
    Z(S) = \frac{\log(S)-\mu_2}{\sigma_2} \simeq \log(S).
\end{equation}
As such, in our empirical specifications, we use $Z(S)$ instead of $\log(S)$ because it improves interpretability and because it permits $S$ to be a member of the real-numbers.

\printbibliography